\documentclass[11pt,a4paper]{article}
\usepackage{jheppub}
\pdfoutput=1
\usepackage{epsfig}
\usepackage{latexsym}
\usepackage{amsfonts}
\usepackage{amsmath}
\usepackage{amsthm}
\usepackage{amssymb}
\usepackage{amsbsy}
\newcommand{\ft}[2]{{\textstyle\frac{#1}{#2}}}

\newcommand{\CC}{\mathbb{C}} % Complessi
\newcommand{\RR}{\mathbb{R}} % Reali
\newcommand{\G}{\mathcal{G}}

\def\calo         {{\cal O}}

\newsavebox{\uuunit}
\sbox{\uuunit}
    {\setlength{\unitlength}{0.825em}
     \begin{picture}(0.6,0.7)
        \thinlines
        \put(0,0){\line(1,0){0.5}}
        \put(0.15,0){\line(0,1){0.7}}
        \put(0.35,0){\line(0,1){0.8}}
       \multiput(0.3,0.8)(-0.04,-0.02){12}{\rule{0.5pt}{0.5pt}}
     \end {picture}}

\def\be{\begin{equation}}
\def\ee{\end{equation}}
\def\bea{\begin{eqnarray}}
\def\eea{\end{eqnarray}}

\def\a{\alpha}
\def\b{\beta}
\def\h{\eta}
\def\g{\gamma}
\def\G{\Gamma}
\def\d{\delta}
\def\e{\epsilon}
\def\D{\Delta}
\def\l{\lambda}
\def\L{\Lambda}

\def\f{\phi}
\def\vf{\varphi}
\def\m{\mu}
\def\n{\nu}
\def\o{\omega}
\def\O{\Omega}
\def\p{\pi}
\def\r{\rho}

\def\s{\sigma}

\def\t{\tau}

\def\th{\theta}

\def\id{{\mathbb I}}

\def\mso{\mathfrak{so}}

\def\msl{\mathfrak{sl}}
\def\msp{\mathfrak{sp}}

\def\mhs{\mathfrak{hs}}

\def\wk{\widehat \kappa}
\def\wf{\widehat f}
\def\wg{\widehat g}
\def\wA{\widehat A}
\def\wB{\widehat B}
\def\wW{\widehat W}
\def\wS{\widehat S}
\def\wK{\widehat K}
\def\we{\widehat \epsilon}

\def\wcS{\widehat {\cal S}}
\def\wF{\widehat \Phi}
\def\wM{\widehat M}
\def\wL{\widehat L}

\def\ty{\tilde y}
\def\tz{\tilde z}

\def\ty{\tilde y}
\def\tz{\tilde z}

\def\ty{\tilde{y}}

\def\tf{\tilde{f}}
\def\tg{\tilde{g}}
\def\tn{\tilde{n}}
\def\tm{\tilde{m}}

\def\ket#1{|#1\rangle}
\def\bra#1{\langle#1|}

\newcommand{\eq}[1]{(\ref{#1})}

\def\sF{{{ F}\!\!\!\!\hskip.8pt\hbox{\raise1pt\hbox{/}}\,}}
\def\som{{{ \omega}\!\!\!\!\hskip.8pt\hbox{\raise1pt\hbox{/}}\,}}
\def\sJ{{{\rm J}\!\!\!\!\hskip.8pt\hbox{\raise1pt\hbox{/}}\,}}

\def\F{\Phi}
\def\pa{\partial}

\def\to{\rightarrow}
\def\nonu{\nonumber \\{}}
\def\half{{1 \over 2}}

\def\tr{{\rm tr}}
\title{On big crunch solutions in Prokushkin-Vasiliev theory}
\author[a,b]{Carlo Iazeolla,}
\author[a]{Joris Raeymaekers}

\affiliation[a]{Institute of Physics of the ASCR, \\
Na Slovance 2, 182 21 Prague 8, Czech Republic.}
\affiliation[b]{NSR Physics Department,\\ G. Marconi University, Rome, Italy}

\emailAdd{iazeolla@fzu.cz}
\emailAdd{joris@fzu.cz}

\abstract{We construct simple solutions of three-dimensional higher spin gravity interacting with matter in which only the scalar and spin-two fields are excited. They preserve Lorentz symmetry
and are similar to the four-dimensional solutions constructed by Sezgin and Sundell, with the difference that the additional twisted sectors  of the three-dimensional theory
are excited. Furthermore, the three-dimensional system contains an extra parameter $\lambda$ which allows us to vary the mass of the scalar. Among other reasons, the resulting solutions are interesting for the holographic study of cosmological singularities: they describe the growth of a Coleman-De Luccia  bubble in anti-de Sitter space,  ending in a
 big crunch singularity. We initiate the holographic study of these solutions, finding evidence for their interpretation within  a multi-trace deformation which renders the dual field theory unstable. The limit $\l \to 0$ is particularly interesting as it captures effects of a running coupling in a large-$N$ interacting fermion model. We also propose a generalization of our solutions, consisting of a dressing with Lorentz-invariant projectors. This additional sector remains non-trivial when the scalar field is turned off.}
 \preprint{arXiv:1510.08835}

\begin{document}
\maketitle

\section{Introduction and summary}
In recent years, much has been learned about the holographic duality between conformal field theories (CFTs) containing conserved higher spin currents and higher spin gravity theories in Anti-de-Sitter ($AdS$) space. In these examples the CFTs are relatively tractable, leading to the exciting prospect of fully solving quantum higher spin gravity. In three spacetime dimensions,
the system of equations describing massless higher spin fields interacting with matter was presented by Prokushkin and Vasiliev \cite{Prokushkin:1998bq} (see also \cite{Vasiliev:1992ix}, \cite{Vasiliev:1997dq}). As we shall review below, in its simplest version this theory describes a  complex scalar field coupled to massless fields of spin $1,2,3, \ldots$. The dual conformal field theory, as proposed by Gaberdiel and Gopakumar \cite{Gaberdiel:2012uj},\cite{Gaberdiel:2012ku}, is a particular limiting case  of the unitary $W_N$  minimal models.

The Prokushkin-Vasiliev (PV) system allows for a consistent truncation where the scalar field is switched off; all the gauge fields, that in $D=3$ do not contain any propagating degrees of freedom, are then described by a Chern-Simons  gauge field taking values in the higher spin algebra $\mhs[\l ]$. So far most of the holographic investigations of the theory have taken place either within the Chern-Simons subsector or in an approximation where the scalar field is treated as a small perturbation. To progress
beyond this  it would be  interesting to explore the space of solutions of the PV  system where the scalar field is nontrivial and fully backreacted. For the four-dimensional Vasiliev system, many such solutions have by now been constructed \cite{Sezgin:2005pv,Sezgin:2005hf,Iazeolla:2007wt,Didenko:2009td,Iazeolla:2011cb,Iazeolla:2012nf,Gubser:2014ysa,Bourdier:2014lya}, while  in three dimensions examples are %scarce \cite{Prokushkin:1998bq,Gutperle:2011kf,} \comment{Do these have the scalar switched on?}
nonexistent to our knowledge.

There are  nevertheless several reasons to be interested in
three-dimensional solutions. First of all, due to the topological nature of the higher spin sector, it may prove simpler to work out the details of the holographic dictionary for the three-dimensional theory than for higher dimensions.
Furthermore, the PV system contains the extra free parameter $\l$ allowing us to vary the mass of the scalar. We should also mention that a better understanding of holography in the scalar sector of the theory may shed light on open  issues in higher spin holography in de Sitter space \cite{Anninos:2012ft},\cite{Anninos:2013rza}.

In this work, we take a first step towards bridging this gap by constructing some of the simplest  solutions of the PV system with a nontrivial scalar profile, and by initiating the
study of their dual interpretation. We will construct  solutions preserving the Lorentz subgroup  of the higher spin symmetry, in which
only the scalar field and the spin-two gravity sector are excited. In  the 4D Vasiliev system,   similar solutions were constructed by  Sezgin and Sundell \cite{Sezgin:2005pv},\cite{Sezgin:2005hf}. One notable difference with the Sezgin-Sundell  solutions is  that the three-dimensional PV system contains additional `twisted' sectors (see \cite{Kessel:2015kna} for a detailed discussion),
which turn out to be inevitably nontrivial in our class of solutions.

The solutions we obtain are interesting for the holographic study of cosmological singularities. They obey boundary conditions in which the AdS vacuum is unstable and
 describe the materialization and growth of a Coleman-De Luccia \cite{Coleman:1980aw} bubble within AdS. It is well-known that the continuation of such solutions in the interior of the bubble is an FRW geometry ending in a
 big crunch singularity\footnote{To be more precise, our solutions appear singular from the point of view of ordinary spin-2 differential geometry. The final verdict on the regularity of our solutions should come from
  a fuller understanding of higher spin geometry.}.
Our solutions  are therefore examples of crunching solutions in  higher spin gravity; similar solutions in other holographic theories have been studied extensively in the literature following the work of Hertog and Horowitz \cite{Hertog:2004rz}\footnote{See e.g. \cite{Hertog:2005hu}-\cite{Barbon:2011ta} for an (incomplete) list of further references.}.

We then go on to argue that our solutions are asymptotically AdS and proceed to study their near-boundary behaviour. To read off  quantities  in the holographically dual CFT from this asymptotic behaviour, such as
sources and vacuum expectation values (VEVs) of various operators,
requires the process of holographic renormalization \cite{de Haro:2000xn}, which is unfortunately not yet available for the PV system when the scalar is excited.
In this work we will therefore limit ourselves to justifying the asymptotic behaviour of the fields by comparing with standard scalar-gravity theories, and we
will propose a dual picture for our solutions based on some reasonable assumptions for the holographic dictionary. This picture is again similar to that
in the other examples in the literature \cite{Hertog:2004rz}-\cite{Barbon:2011ta}: the solutions obey boundary conditions which correspond to turning on a wrong-sign marginal multi-trace deformation which renders the
theory unstable. The VEV of the operator dual to the scalar displays a runaway behaviour and becomes infinite in finite global time, at the instance when the
crunch singularity hits the boundary.
 Evidence for this picture comes from considering the CFT effective action for the operator dual to the bulk scalar: we will see that the two-derivative approximation  allows
  for solutions with runaway behaviour which precisely matches the boundary  behaviour of the scalar field. We will devote special attention to the  $\l \to 0$ limit of our solutions
  where the scalar saturates the Breitenlohner-Freedman bound. In this case the bulk solution captures the effects of a running coupling in the boundary theory, which can be understood using the known fermionic description \cite{Gaberdiel:2011wb,Gaberdiel:2013jpa} of the dual theory.

Furthermore, we find that it is possible to generalize our solutions by expanding the twistor-space connection of the PV system on a set of Lorentz-invariant projectors, thereby realizing a three-dimensional analogue of the solutions first found in \cite{Iazeolla:2007wt} for the 4D Vasiliev equations. This expansion brings in new, discrete parameters that turn on seemingly new, global degrees of freedom. Indeed, on the PV equations the twistor-space connection can locally be entirely solved in terms of the matter fields up to monodromies, and here we find that its component on the projector sector remains non-trivial even when the scalar is switched off, giving rise to  solutions without matter that seem to be globally inequivalent to the $AdS_3$ background.

This paper is structured as follows. In section \ref{secVas} we review the Prokushkin-Vasiliev system, paying special attention to the projection to a bosonic theory containing a single complex scalar, which plays a role in the holographic duality proposal of \cite{Gaberdiel:2012uj}, \cite{Gaberdiel:2012ku}. In section \ref{Sec:solution} we present our solutions, deriving the profile of  the scalar field for aribitrary $\l$ and the more involved  splitting of the higher spin sector into Lorentz connection and vielbein fields for the
 particular case $\l = \half$. We also briefly discuss the more general solutions which contain the star-algebra projectors. Subsequently, in section \ref{secgeom}, we discuss the three-dimensional geometry of the solutions, arguing that they are asymptotically anti-de Sitter and that they describe the materialization and growth of a Coleman-De Luccia bubble within a metastable AdS vacuum, ending eventually in a big crunch singularity. In section \ref{sechol} we initiate the holographic study of our solutions,
 understanding their asymptotic behaviour by comparing with scalar-gravity theories. We  propose a  dual picture in a theory deformed by multi-trace deformations, and study in detail
$\l \to 0$ limit where we compare with  the fermionic description of the dual theory.

\section{Elements of $D=3$ higher-spin gravity coupled to matter}\label{secVas}

\subsection{Master fields and extended oscillator algebra}\label{Sec:kinematics}

The fully non-linear field equations of three-dimensional higher-spin gravity with propagating matter fields are formulated \cite{Vasiliev:1992ix,Vasiliev:1997dq,Prokushkin:1998bq} in terms of differential forms living on a base manifold locally given by the direct product of the commuting spacetime manifold $\cal X$ and a non-commutative twistor space $\cal Z$, and valued in the associative algebras of functions on a fibre manifold $\cal Y \times \cal A$, where $\cal Y$ is another twistor space and $\cal A$ contains possible internal matrix algebras. The total space,  is therefore locally equivalent to the product $\mathfrak{C} = \cal X \times \cal Z \times \cal Y \times \cal A$, with coordinates $(x^\m, z_\a, y_\a, T_a)$, where $\m=0,1,2$, $\a=1,2$ and the $T_a$ are generators of internal symmetry algebras. In the present paper, we shall take the latter to coincide with a set of four elements $\{\psi_{1,2},k,\r\}$ generating two Clifford algebras\footnote{In general, it is possible to add a set of generators of non-abelian internal matrix algebras $Mat_n(\CC)$ \cite{Prokushkin:1998bq}.}.

In particular, the variables entering the equations are three master fields: a spacetime one-form $\widehat W = dx^\m \widehat W_\m (y,z;\psi_{1,2},k,\rho|x)$, which contains the gauge connections for all spins accompanied by a set of auxiliary connections, a zero-form $\widehat B=\widehat B(y,z;\psi_{1,2},k,\rho|x)$, containing the local degrees of freedom of the theory (the matter fields and their on-shell non-trivial derivatives), and a twistor-space one-form connection $\widehat S=dz^\a\widehat S_\a(y,z;\psi_{1,2},k,\rho|x)$ which contains no local degrees of freedom (in the sense that, on-shell, it can be entirely solved in terms of $\widehat B$ up to monodromies). Following the convention of \cite{Sezgin:2005pv}, we shall henceforth use hatted variables to distinguish $z$-dependent twistor-space functions and drop the hats for purely $y$-dependent variables.  We shall now turn to collecting the relevant properties of all the arguments of the above master fields, referring the reader to Appendix \ref{App:conv} for our spinor conventions.

The twistor-spaces $\cal Y$ and $\cal Z$ are equipped with an associative and non-commutative $\star$-product algebra, with respect to which their coordinates $y_\a$ and $z_\a$ satisfy two independent Heisenberg oscillator algebras,
\be
 [ y_\a, y_\b]_\star \ = \ 2 i \e_{\a\b},\qquad [ y_\a, z_\b]_\star =0\ , \qquad  [ z_\a, z_\b]_\star \ = \  -2 i \e_{\a\b}\ , \label{yzcomm} \ee
  \be[ y_\a,\widehat f(z,y)]_\star \ = \ 2 i {\pa \widehat f \over \pa y^\a} \ ,
    \qquad [ z_\a, \widehat f(z,y)]_\star \ = \ -2 i {\pa \widehat f \over \pa z^\a}\ .
    \label{yzder} \ee
    %
  %  \bea
    % y_\a \star y_\b &=& y_\a y_\b + i \e_{\a\b}\qquad z_\a \star z_\b = z_\a z_\b - i \e_{\a\b}\qquad y_\a \star z_\b = y_\a z_\b - i \e_{\a\b}
    % \eea
%
In a specific choice of basis (i.e., normal-ordering with respect to the combinations $A^-_\a:= \frac{1}{2}(y_\a+z_\a)$ and $ A^+_\a  := \frac{1}{2i}(y_\a-z_\a)$, $[A^-_\a,A^{+\b}]_\star=\d_\a^\b$), the $\star$-product among arbitrary functions on $\cal Z \times \cal Y$  is realized as
\bea \label{starprod}
 (\widehat f\star \widehat g)(z,y)&=&\frac{1}{(2\pi)^2}\int d^2ud^2v\;\exp(iu_\a v^\a)
   \,\widehat f(z+u,y+u)\,\widehat g(z-v,y+v)  \ .
%     y_\a \star y_\b &=& y_\a y_\b + i \e_{\a\b}\qquad z_\a \star z_\b = z_\a z_\b - i \e_{\a\b}\qquad y_\a \star z_\b = y_\a z_\b - i \e_{\a\b}
\eea
Of crucial importance to the formulation of the equations is also the inner kleinian operator
\be  \wk \ := \ (-1)^{\widehat N}_\star \ , \qquad \widehat N \ = \ A^{+\a}\star A^-_{\a} \ = \ -\frac{i}{2} y^\a  z_\a \ , \ee
which, in normal order, reduces to \cite{Prokushkin:1998bq,Iazeolla:2008ix}
\be \wk \ = \ e^{iy^\a z_\a} \ , \ee
and has the properties
\be \wk \star \wf(z,y) \ = \ \wf(-z,-y) \star  \wk \ , \qquad \wk \star \wk \ = \ 1 \ .\ee
The Prokushkin-Vasiliev equations also make use of two pairs of Clifford elements, $(\psi_1, \psi_2)$ and $(\r,k)$, such that
\bea \{ \psi_i, \psi_j \} &=& 2 \d_{ij}, \qquad i = 1,2 \ , \label{psialg}\eea
while commuting with all other variables, and
\bea k^2 \ = \ \r^2 =1\ , \quad \{\r, k\} =0\ ,\quad \{ k, y_\a\} \ = \ \{ k, z_\a\}=0
\ ,\quad [\r, y_\a ] \ = \ [ \r, z_\a ]=0 \ .\eea
The Clifford element $k$ induces a doubling of all fields in the model which is related to  the ${\cal N}=2$ supersymmetry subalgebra of the global symmetry algebra of the theory. The presence of $\r$ makes it possible to write the equations for the $k$-dependent $\wS_\a$ manifestly as a deformed oscillator algebra \cite{Vasiliev:1997dq}. The Clifford element $\psi_1$ is responsible for the embedding of AdS translations inside the maximal finite subalgebra of the bosonic global symmetry algebra of the theory, i.e., the semisimple $AdS_3$ isometry algebra $\mso(2,2)  \simeq \msp(2,\RR)\oplus\msp(2,\RR)$, with $\msp(2,\RR)\simeq\mso(2,1)\simeq\msl(2,\RR)$: identifying the Lorentz subalgebra as the diagonal $\mso(2,1)$, spanned by elements $J_a \leftrightarrow M_{\a\b}$  (see Appendix \ref{App:conv} for our spinor conventions and more details)
\be M_{\a\b} \ = \ \frac{1}{4i}\{y_\a,y_\b\}_\star \ \equiv \ \frac{1}{2i}y_\a y_\b \ , \ee
the translations are realized as $P_a = J_a\psi_1 \leftrightarrow M_{\a\b}\psi_1$, and one can use the projectors $\Pi_\pm:=(1\pm\psi_1)/2$ to single out the two simple components of $\mso(2,2)$. Note that, as explained in \cite{Prokushkin:1998bq} and \cite{Vasiliev:1999ba} and as we shall briefly review in Section \ref{Sec:hsalg}, this is only one of a one-parameter family of possible oscillator realizations of $\mso(2,2)$. %the relevant one for describing massless matter fields.
Finally, the role of $\psi_2$ is to implement, via its anticommutation property with $\psi_1$, a twist operation $\psi_1 \rightarrow -\psi_1$ changing the sign of the translation generators (which corresponds to an involutive automorphism of the $AdS_3$ isometry algebra \eq{sogenc}): essentially, its inclusion in the expansion of the master fields realizes a grading of the commutators with respect to the translations which is crucial in order to include non-trivial propagating matter fields via the field equation involving $\wB$.

Furthermore, the master fields obey the following reality conditions,
\be
\widehat W^\dagger = - \widehat W, \qquad \widehat S_\a^\dagger = - \widehat S_\a, \qquad \widehat B^\dagger = \widehat B \ ,\label{real}
\ee
where $\ ^\dagger$ acts on the variables on $\mathfrak{C}$ as
\be
[A(z,y;\psi_{1,2},k,\r)]^\dagger \equiv \bar A^{rev} (-z,y; \psi_{1,2},k,\r)
\ee
and $\ ^{rev}$ means reversing the order of the Grassmann elements.

\subsection{The Prokushkin-Vasiliev equations and their consistent truncations}

We are now in a position to write down the full Prokushkin-Vasiliev equations,
\bea
&d\widehat W + \widehat W \star \widehat W \ = \ 0&\label{hseq1}\\
&d\widehat B + [ \widehat W , \widehat B]_\star \ = \ 0 &\label{hseq2}\\
& d \widehat S_\a +[ \widehat W, \widehat S_\a]_\star \ = \ 0 & \label{hseq3}\\
&  [\widehat S_\a, \widehat B]_\star \ = \ 0 &\label{hseq4}\\
& [\widehat S_\a, \widehat S_\b ]_\star \ = \  - 2 i \e_{\a\b} ( 1 + \wB  \star \wK )& \label{hseq5}
 \eea
where the $\wedge$-product of differential forms is always understood and we have defined the total kleinian operator
\be \wK \ := \ k \,\wk \ .   \ee
%.
The system is manifestly gauge invariant under the transformations
\bea
\d_{\we}\, \widehat  W & = & d \widehat \e + [\,\widehat W  , \widehat\e\,]_\star \ , \\
\d_{\we} \,\widehat  B & = &- [\,\we, \widehat  B\,]_\star \ ,\\
\d_{\we}\, \widehat  S_\a & = & -[\,\we, \widehat S_\a]_\star \ ,\label{gt}
\eea
where $\widehat \e = \widehat \e (z,y;\psi_{1,2},k|x)$. %\footnote{Note that the gauge parameter does not depend on $\r$. Moreover, as explained in detail in \cite{Prokushkin:1998bq}, the degrees of freedom of the theory, residing in $after proper boundary conditions are imposed, }.
As explained in \cite{Prokushkin:1998bq}, by virtue of the the involutive automorphism $\r \rightarrow -\r$, $\wS_\a \rightarrow -\wS_\a$, the system \eq{hseq1}-\eq{hseq5} can be consistently truncated to the one with fields $\wW$ and $\wB$ independent of $\r$, and $\wS_\a$ linear in $\r$,
\be \wW \ = \ \wW(z,y;\psi_{1,2},k|x)\ , \quad  \wB \ = \ \wB(z,y;\psi_{1,2},k|x)\ , \quad \wS_\a \ = \ \rho\,\wcS_\a(z,y;\psi_{1,2},k|x) \ .\label{rho}\ee
Note that such reduced variables satisfy
\be \wK\star\wW \ = \ \wW\star\wK \ , \quad \wK\star\wB \ = \ \wB\star\wK \ , \quad \wK\star\wS_\a \ = \ -\wS_\a\star\wK \ .\label{Kcond}\ee

In this paper we shall focus on the bosonic truncation of the Prokushkin-Vasiliev equations, which amounts to further imposing that $\wW$ and $\wB$ are of even degree in the total number of $(y,z)$ oscillators and $\wS_\a$  is of odd degree. This projection is implemented via the conditions
\be  \wk\star\wW \ = \ \wW\star\wk \ , \qquad \wk\star\wB \ = \ \wB\star\wk \ , \qquad \wk\star\wS_\a \ = \ -\wS_\a\star\wk \ ,  \label{bosonic}\ee
which, together with \eq{Kcond}, implies that $k$ is central in the reduced bosonic theory \eq{rho}, and as a consequence the latter can be projected to two independent subsectors with the help of the projectors $P_\pm:=(1\pm k)/2$.
Therefore, one can obtain the projected components of the master-fields from the two independent subsystems
\bea
&d\wW^\pm + \wW ^\pm\star \wW^\pm \ = \ 0&\label{hseq1pm}\\
&d\wB^\pm + [ \wW^\pm , \wB^\pm]_\star \ = \ 0 &\label{hseq2pm}\\
& d \wS_\a^\pm+[ \wW^\pm, \wS^\pm_\a]_\star \ = \ 0 & \label{hseq3pm}\\
&  [\wS^\pm_\a, \wB^\pm]_\star \ = \ 0 &\label{hseq4pm}\\
& [\wS^\pm_\a, \wS^\pm_\b ]_\star \ = \  - 2 i \e_{\a\b} ( 1 \pm \wB^\pm  \star \wk )& \ .\label{hseq5pm}
 \eea
As we shall review below, the $P_+$- and $P_-$-projected sectors each describe a single propagating complex scalar field. In the refined version of the Gaberdiel-Gopakumar duality proposed in \cite{Gaberdiel:2012ku}, the bulk theory consists of only one  projected sector. Our strategy in this paper will be to construct solutions of the reduced bosonic theory satisfying (\ref{rho}, \ref{bosonic}), and set to $k$ equal 1 (or -1) at the end in order to obtain a solution in the $P_+$- ($P_-$-) projected sector.
We shall postpone examining another consistent truncation to Section \ref{Sec:hsalg}, where the necessary notions of $\nu$-vacua and deformed oscillators are reviewed.

\subsection{Maximally symmetric vacua and higher-spin algebras}\label{Sec:hsalg}

As shown in \cite{Prokushkin:1998bq}, unlike the 4D Vasiliev theory, the 3D theory has a one-parameter family of inequivalent $AdS$ vacua, labelled by the constant\footnote{We will not consider here the possibility of letting the identity component of $\widehat B$ vary over spacetime.} identity component $\n\in\RR$ of $\widehat B$, around which the physics is different: in particular, the mass of the scalars\footnote{Let us remind the reader that in $AdS_3$ spacetime the Klein-Gordon equation takes the form
\be \left(\Box-m^2\right)\phi\ \equiv \ \left(\Box-M^2_\L - M^2\right)\phi \ = \ \left(\Box+\frac{3}{4}-M^2\right)\phi \ = \ 0 \nonumber\ , \ee
where $m^2$ separates into a mass-like term associated to the spacetime curvature (and which, reinstating the $AdS$ radius $L$ that we are setting to $L=1$ throughout the paper, reads $M^2_\L = -\frac{3}{4L^2}$), coming from a curvature coupling  in the action, and a proper mass term $M^2$. We shall follow the convention in, e.g., \cite{Prokushkin:1998bq}, and refer to the particles with $M^2=0$ as massless.} and the higher-spin algebras depend on $\n$.
 Setting $\widehat B = \wB_0 = \nu$ solves eqs. \eq{hseq2} and \eq{hseq4} automatically. The field $\widehat S_\a$ must then be constant in spacetime  and satisfy the deformed oscillator algebra
\be
\,[\widehat S_{0\a}, \widehat S_{0\b} ]_\star = - 2 i \e_{\a\b} ( 1 + \n \widehat K ).
\ee
The simplest solution (more general solutions will be discussed in Section \ref{secgen})  which obeys the reality condition (\ref{real}), which we review in Appendix \ref{App:defosc},  is \cite{Prokushkin:1998bq}
\bea
\widehat S_{0\a} & = & \r\left\{z_\a -\frac{\n}{8}\int_{-1}^1 ds(1-s) \left((y_\a+z_\a)e^{{i \over 2} (1+s) u}\star \,_1F_1 \left[ \half,2,-\n \wK\ln |s|\right] \right. \right. \nonu
 & &\ \ \ + \left. \left. (y_\a-z_\a)e^{{i \over 2} (1+s) u}\star \,_1F_1 \left[ \half,2,\n \wK\ln |s|\right] \right)\star \wK\right\} \nonu
 & = & \r z_\a \left[ 1+ \n \int_{-1}^1 ds(1-s) \left( F^-(\n \ln |s| ) e^{{i \over 2} (1+s) u } +  F^+(\n \ln |s| )e^{{i \over 2} (1-s) u} k\right)\right] \nonu
 & =: & \tilde z_\a \label{S0reg}
\eea
where we are defining $u:= y^\a z_\a$, the only non-trivial Lorentz-invariant combination of the oscillators alone,
\be
F^\pm (x) \equiv {1\over 8}\left( \,_1F_1 \left[ \half,2,x\right] \pm  \,_1F_1 \left[\half,2,-x\right]\right) \ ,
\ee
and $\,_1F_1\left[\half,2,x\right] $ is the confluent hypergeometric function. Furthermore, there exist deformed oscillators $\tilde y_\a$ that $\star$-commute with $\tilde z_\a$\footnote{For notational simplicity, we shall omit the hat over the deformed oscillators $\ty_\a$ and $\tz_\a$ even though they are functions of $z_\a$ for any $\nu\neq 0 $. } \cite{Prokushkin:1998bq}
\be\label{defy}
\ty_\a \ = \ y_\a-  \n \int_{-1}^1 ds\,(1-s) \,e^{{i \over 2} (1+s) u } \left( y_\a F^-(\n \ln |s| )  - z_\a  F^+(\n \ln |s| )k\right) \ ,
\ee
such that
\bea
\,[\tilde z_\a,\tilde z_\b ]_\star &=& - 2 i \e_{\a\b} ( 1 + \n  \wK )\ ,\label{defzalg}\\
\,[\tilde y_\a,\tilde y_\b ]_\star &=&  2 i \e_{\a\b} ( 1 + \n k )\ ,\label{defyalg}\\
\,[\tilde y_\a,\tilde z_\b ]_\star &=& 0 \ .
\eea
Note that $\ty_\a|_{\n=0}=y_\a $ and $\tz_\a|_{\n=0}=\r z_\a$. From now on, we shall adapt our notation to an expansion over these vacua, and use hats over any variable that depends on $\tz_\a$, while variables that only depend on $\ty_\a$ will be unhatted even though the latter contain $z_\a$. For $\n=0$ this notation reduces to the one so far adopted.

An important property of the deformed oscillator algebra is that the bilinears
\be
M_{\a\b} = -{i \over 4} \{ \tilde y_\a, \tilde y_\b \}_\star
\ee
satisfy the $\msp(2,\RR)$ commutation relations,
\bea
[M_{\a\b},M_{\g\d}]_\star \ = \ 4\e_{(\b|(\g}M_{\d)|\a)} \ ,\label{yLorentz}
\eea
and
\be [M_{\a\b},\ty_\g]_\star \ = \ 2\e_{(\b|\g}\ty_{|\a)} \ee
for any $\n$. Eq. \eq{hseq3} implies that the vacuum solution $\wW_0$ commutes with $\wS_{0\a}$, i.e., that $\wW_0 = W_0(\ty; \psi_{1,2},k)$. The $AdS_3$ geometry is then encoded in \eq{hseq1} for the flat connection
\be  W_0 \ = \ e_0 + \o_0 \ , \qquad  \o_0 \ = \ \frac{1}{4i}\o_0^{\a\b}  \{ \tilde y_\a, \tilde y_\b \}_\star \ , \quad e_0 \ = \ \frac{1}{4i}e_0^{\a\b}  \{ \tilde y_\a, \tilde y_\b \}_\star\psi_1\label{W0}\ee
where $e_0$ and $\o_0$ are the $AdS_3$ vielbein and the Lorentz-connection, for which \eq{hseq1} results in zero-torsion and zero-curvature conditions for any $\nu$.

The global symmetry algebra of the theory corresponds to the stability subalgebra of these maximally symmetric vacua, containing parameters $\we_{gl}=\we_{gl}(\ty;\psi_1,k)$ \cite{Prokushkin:1998bq}. %\footnote{As the Clifford element $\psi_2$ distinguishes the sector  of dynamical fields from the twisted sector, in the terminology of ...}.
Restricting our attention to the bosonic subalgebra, which will be of relevance in the following, we find that the the latter contains elements $\we^\pm_{gl}(\ty;\psi_1)$\footnote{The explicit dependence on $k$ is absent because, as explained before, in the bosonic projection $k$ becomes central and the master-fields can be projected onto the two independent subsectors where $k=\pm1$. The dependence on $\psi_2$ is absent because, as we shall recall immediately below, the expansion on $\psi_2$ in the master-fields separates the physical and the twisted sector of the theory. A symmetry parameter linear in $\psi_2$ would mix the two sectors, and for this reason should be ruled out in a unitary theory \cite{Prokushkin:1998bq}.},  which are expansions on symmetrized even-degree polynomials in $\ty_\a$. By means of \eq{defy}, the deformed oscillator algebra \eq{defyalg} turns out to be embedded in the tensor product of the undeformed oscillator algebras of $(y,z)$ and $k$. Taking $\star$-commutators of the bosonic parameter $\e_{gl}(\ty;\psi_1,k)$ and then projecting onto $k=\pm1$ then amounts to realizing the algebra $\mhs(2;\n)_\pm\oplus\mhs(2;\n)_\pm$, where each $\mhs(2; \n)_\pm$ corresponds to an inequivalent higher-spin extension of the maximal finite bosonic subalgebra $\msp(2,\RR)$. More specifically,  $\mhs(2;\n)_\pm$ is the Lie algebra arising from the unital associative algebra
\bea Aq(2; \n)_\pm  \ = \ \frac{{\cal U}[\msp(2,\RR)]}{{\cal I}[C_2+\frac{3\pm2\n-\n^2}{4}]} \eea
via commutators. ${\cal U}[\msp(2,\RR)]$ is the universal enveloping algebra of $\msp(2,\RR)$, while ${\cal I}[C_2+\frac{3\pm2\n-\n^2}{4}]$ is a two-sided ideal generated by the relation in the argument, that fixes the value of the Casimir of $\msp(2,\RR)$  in terms of $\nu$, and the algebras $Aq(2;\n)_\pm$ correspond to the associative algebra of even functions of the $\ty_a$ projected via $P_\pm$. In the notation of \cite{Gaberdiel:2012uj} and of the Introduction, $\mhs(2;\n)_\pm = \mhs[\l]$, where $\l = (1\mp\n)/2$ plays the role of a 't Hooft-like coupling of the dual CFT \cite{Gaberdiel:2012uj}.
Finally, note that $Aq(2;\n)_\pm$ admit a uniquely defined trace operation (descending from the supertrace of $Aq(2;\n)$, see \cite{Prokushkin:1998bq}) that consists of the projection onto the coefficient of the unit element,
\be \tr(f^\pm(\ty)) \ = \ f^\pm(0) \ , \ee
and which, upon including also a dependence on $\psi_{1,2}$, is defined as
\be \tr(F^\pm(\ty;\psi_{1,2})) \ = \ F^\pm(0)|_{\psi_{1,2}=0} \ .\label{trdef} \ee

An expansion around the $\n$-vacua $(\wW_0,\wB_0,\wS_{0\a})$ defined above, $\wW=\wW_0 + \wW_1+...$, $\wB=\wB_0 + \wB_1+...$, $\wS_{\a}=\wS_{0\a} + \wS_{1\a}+...$,  shows \cite{Prokushkin:1998bq} that the local degrees of freedom of the theory reside in the linearized zero-form $C^\pm(\ty; \psi_{1,2})=\wB_1$.  As anticipated, since the background covariant derivative acting on it in \eq{hseq2pm} $d+[\wW_0, ... ]$ explicitly contains $\psi_1$ (see \eq{W0}), the expansion in $\psi_2$
\be C^\pm(\ty; \psi_{1,2}) \ = \ C^{\pm,tw}(\ty; \psi_{1})+C^{\pm,phys} (\ty; \psi_{1})\psi_2 \ , \ee
separates the field content of the zero-form into an infinite set of finite-dimensional $\mso(2,2)$-multiplets of tensor fields contained in $C^{\pm,tw}$, on which \eq{hseq2pm} imposes Killing equations, and the infinite-dimensional multiplet in $C^{\pm,phys}$ formed by the modes of two pairs of real scalar fields of mass $M_\pm^2 = \n(\n\mp 2)/4$, on which \eq{hseq2pm} imposes the Klein-Gordon equation. Each pair belongs to one of the two independent subsectors obtained via the projectors $P_\pm$, and the doubling within each subsector is due to $\psi_1$\footnote{The full Prokushkin-Vasiliev theory, on the other hand, contains these four real scalar fields, of masses $M^2_\pm = \n(\n\mp2)/4$, together with four fermions of masses $M^2_\pm = \n^2/4$ arranged in a ${\cal N}=2$ supersymmetry multiplet \cite{Prokushkin:1998bq}.} . The perturbative analysis also shows that $\wS_{1\a}$ can locally be solved entirely in terms of $C$, and that the field content of $\o:=\wW_1$ also separates into
\be\o = \o^{\pm,phys}(\ty; \psi_1)+\o^{\pm,tw}(\ty;\psi_1)\psi_2\ ,\ee
where $\o^{\pm,phys}$ contains the bosonic gauge fields of all spins, which in $D=3$ are all topological, and $\o^{\pm,tw}$ contains twisted gauge fields.

It is presently not clear what is the correct interpretation of the fields in the twisted-sector, nor what their role could be in the Gaberdiel-Gopakumar duality conjecture \cite{Gaberdiel:2012uj}, since there is no natural dual for them in the boundary theory. Moreover, in \cite{Kessel:2015kna} it was found that, up to second order in perturbation theory, there exists a field redefinition that consistently truncates away the twisted sector, leaving only the physical fields. This issue remains to be investigated at the non-perturbative level.

In this paper we shall construct and analyze a Lorentz-invariant exact solution which is a $D=3$ counterpart of the four-dimensional Sezgin-Sundell solution found in \cite{Sezgin:2005pv}, mainly focusing on the physical sector and on the $\mhs[\l=1/2]$-theory ($\n=0$).
We shall separate $\wB$ into the vev corresponding to the $\n$-vacua and a fluctuation $\widehat \F$,
\be\label{nplusF}
\wB \ = \ \n + \widehat \F, \qquad \tr \widehat\F \ = \ 0.
\ee
%
%(The physical Klein-Gordon field in AdS turns  out to be the component $\tr  \widehat \F \psi_2$ \cite{}.)
Correspondingly, we shall expand the master fields of a general solution to the Vasiliev equations in the symmetrized $\star$-monomials of the deformed oscillators $\tilde y, \tilde z$ as
\bea
\widehat \Psi &=& \sum_{m,n} \Psi^{\a(m), \b (n)} (\psi_i, \r, k|x) W_{\a(m), \b(n)}\ , \\
W_{\a(m), \b(n) } &\equiv& \tilde y_{(\a^1} \star \ldots  \star \tilde y_{\a^m)} \star  \tilde z_{(\b^1} \star \ldots \star \tilde z_{\b^n)} \ .
\eea
The only propagating field in the bosonic theory is the AdS scalar, which corresponds to the component
\bea
\phi \ \equiv \ (\widehat \F\,\psi_2)_{|y=0, z=0, \psi_2=0} \ \equiv \ (\F\,\psi_2)_{|y=0, \psi_2=0} \ = \ \tr(\Phi\psi_2)\ ,\label{physscalar}
\eea
and in the following we shall explain how to extract the physical fields in the gauge-field sector.

Finally, note that in the  massless case $\n=0$ the original system \eq{hseq1}-\eq{hseq5} acquires the automorphism $k\rightarrow -k$, which enables a consistent truncation according to the $k$-parity conditions
\be \wW(k) \ = \ \wW(-k) \ , \qquad \wS_\a(k) \ = \ \wS_\a(-k) \ , \qquad \wB(k) \ = \ -\wB(-k) \ ,\ee
eliminating the $k$-doubling in all the master-fields. Then, subjecting the system also to \eq{rho} and to the bosonic projection, and using the notation \eq{nplusF}, the equations reduce to
\bea
&d\wW + \wW \star\wW \ = \ 0&\label{hseq1red}\\
&d\widehat \F + [ \wW , \widehat \F]_\star \ = \ 0 &\label{hseq2red}\\
& d \wS_\a +[ \wW, \wS_\a]_\star \ = \ 0 & \label{hseq3red}\\
&  [\wS_\a, \widehat \F]_\star \ = \ 0 &\label{hseq4red}\\
& [\wS_\a, \wS_\b ]_\star \ = \  - 2 i \e_{\a\b} ( 1 +   \widehat \F \star \wk )& \ , \label{hseq5red}
 \eea
with $k$-independent master-fields expanded in the undeformed $(y,z)$ oscillators. Note that the bosonic system above is formally identical to each one of the independent bosonic subsystems \eq{hseq1pm}-\eq{hseq5pm}.

As explained in \cite{Prokushkin:1998bq} (see also \cite{Bonezzi:2015igv}), it is possible to further truncate the bosonic PV system down to a model that we shall refer to as ``minimal bosonic PV model'', in analogy with its four-dimensional counterpart (see for example \cite{Sezgin:2005pv}), by virtue of the anti-automorphism $\tau$\footnote{This notation is chosen to keep the four-dimensional notation of \cite{Sezgin:2005pv}. In the original paper \cite{Prokushkin:1998bq} this anti-automorphism was denoted with $\sigma$.},
\bea& \tau(\wf\star\wg) \ = \ (-1)^{{\textrm{deg}}(f){\textrm{deg}}(g)}\tau(\wg)\star\tau(\wf) \ , &\\
&\tau(z_\a, dz^\a; y_\a ; \rho, k, \psi_i) \ = \ (-iz_\a, -idz^\a; iy_\a ; \rho, k, \psi_i) &\label{tauaction}\ .
\eea
The truncation conditions on the bosonic master-fields are
\be \tau(\wW, \wB, \wS_\a) \ = \ (-\wW, \wB, -i\wS_\a) \ ,\label{tautrunc} \ee
and, according to \eq{tauaction}, they leave only one real physical scalar in the spectrum. Indeed, imposing the $\tau$-projection on the bosonic linearized zero-form
\be [C(\ty; \psi_{1,2})]  \ = \  C_0^{tw}(\ty)+C_1^{tw}(\ty)\psi_1+[C_0^{phys} (\ty)+C_1^{phys}(\ty)\psi_1]\psi_2\ee
does not constrain the twisted sector, but requires $\tau(C_0^{phys})= C_0^{phys}$ and $\tau(C_1^{phys})= -C_1^{phys}$, thereby throwing away half of the modes from each physical scalar and reconstructing a single real scalar field from the combination of the residual, allowed modes. As we shall see, the solutions constructed in this paper admit truncation to the minimal bosonic PV model.

We shall now turn to the construction of the exact solution, by first explaining the solution method, which results from a combination of using a gauge function to strip off the $x$-dependence of the master-fields, and of imposing Lorentz-invariance in the twistor-space ${\cal Y}\times{\cal Z}$. For the latter task, it is important to first recall how the local Lorentz symmetry is manifestly realized in the Prokushkin-Vasiliev theory.

\subsection{Manifest spin-2 Lorentz invariance}\label{seclorentz}

Following \cite{Vasiliev:1997dq,Prokushkin:1998bq,Sezgin:2005pv}, in this section we recall the proper generalization of the Lorentz generators \eq{yLorentz} at the fully interacting level and, consequently, how to disentangle a given solution of (\ref{hseq1}-\ref{hseq5}) into fields that transform covariantly under the spin-2 local Lorentz part of the higher spin gauge group. We will only sketch this procedure for  $\n=0$, which is also
the case to which we will apply this formalism in section \ref{secgauge}.

The non-linear modification of $M_{\a\b}$ that properly rotates the $\tz_\a$ on which the master-fields depend, while still ensuring that the $\wS_\a$ have appropriate local Lorentz transformations, is
%To remedy this we would like  modify the Lorentz generators such  hat  the $\widehat S_\a$ transformation has an  an extra term$-\L_\a^{\ \b} \widehat S_\b$ to undo the unwanted Lorentz rotation. %The reason for this is that the gauge transformation (\ref{gt}) does not act on the $\a$-index.
%We introduce modified, field-dependent generators and Lorentz parameters as:
\bea
\widehat M^L_{\a\b} &=& \wM^{tot}_{\a\b} - {i \over 4} \{\widehat S_\a,\widehat S_\b \}_\star \label{Lor} \ ,
%\we_L &=&  \L^{\a\b}\widehat M^L_{\a\b} \ ,
\eea
where
\be \wM^{tot}_{\a\b} = {i \over 4} \left( \{z_\a, z_\b\}_\star - \{ y_\a, y_\b\}_\star \right). \ee
They generate the desired transformation of $\widehat B,\widehat S_\a$\footnote{The desired transformation of $\wS_\a$ is such that the gauge choice which is customary in the perturbative analysis of the PV equations (as it ensures that $\wS_\a$ can be locally reconstructed in terms of $\wB$ at any order in perturbation theory), $z^\a \wS_\a=0$, is maintained under local Lorentz transformations \cite{Kessel:2015kna} or, more generally, that the gauge choice is such that the spinor index of $\wS_\a$ is not carried by any external field \cite{Didenko:2014dwa}. As we shall see, our solutions respect this gauge condition.}:
\bea
\d_{\we_L} \widehat \F &=& -[\we_{tot},\widehat  \F]_\star \\
\d_{\we_L} \widehat S_\a &=& -[\we_{tot},\widehat  S_\a]_\star +\L_\a^{\ \b}\widehat  S_\b
\eea
where the local Lorentz parameters are defined as
\bea
\we_{tot} =\half  \L^{\a\b}\wM^{tot}_{\a\b} \ , \qquad   \we_L &=&\half  \L^{\a\b}\widehat M^L_{\a\b} \ ,
\eea
and we used (\ref{hseq4}, \ref{hseq5}). The master field $W$ transforms as
\be
\d_{\we_L}\widehat  W = \half d \L^{\a\b}\widehat  M^L_{\a\b} + [\widehat  W,\we_{tot}]_\star
\ee
where we used (\ref{hseq3}).

Taking the $ z=0$ components of these expressions we obtain the transformations of the fields $\F, W$
\bea
\d_{\we_L}  \F &=&- [\e_{0},\F]_\star \ , \label{e0F}\\
\d_{\we_L}  W &=&  [W,\e_{0}]_\star + \half d \L^{\a\b}\widehat  M^L_{\a\b \, | z = 0} \ .
%\d_{\e_L}  W &=&  [\e_{0},  W]_\star + d \L^{\a\b}\hat  (T^0_{\a\b} - {i \over 4} \{\hat S_\a,\hat S_\b \}_{\star\, |\tilde z = 0}
\eea
where $\e_0, M_{\a\b}$ are the gauge parameter and Lorentz generators that Lorentz rotate the $y$ oscillators:
\bea
M_{\a\b} &=& - {i \over 4}  \{ y_\a,  y_\b\}_\star \ ,\\
\e_0 &=& \half\L^{\a \b} M_{\a\b} \ .
\eea
Eq. \eq{e0F} tells us that the components in the $y$-expansion of $\F$ transform as Lorentz tensors.
 We want to extract from $ W$ the fields that transform as Lorentz tensors as well as a (spin-2) Lorentz-connection piece $\o^{\a\b}$ which transforms inhomogeneously:
\be
\d_{\we_L}  \o^{\a\b} = d\L^{\a\b} + \o^{\a\g}\L_\g^{\ \b} + \o^{\g\b}\L_\g^{\ \a}.
\ee
Noting that the $\widehat M^L_{\a\b}$, being field-dependent, gauge-transform as
\be
\d_{\e_L} \widehat  M^L_{\a\b \, | z = 0} = -[\e_0,\widehat  M^L_{\a\b \, | z = 0}]_\star -\L_\a^{\ \g} \widehat  M^L_{\g\b \, | z = 0}-\L_\b^{\ \g}\widehat  M^L_{\a\g \, | z = 0}
\ee
and defining
\be
\o \ = \  \half\o^{\a\b}  M_{\a\b}
\ee
we find that the quantity $W -\half \o^{\a\b} \wM^L_{\a\b, | z = 0}$ is composed of Lorentz tensors:
\be
\d_{\we_L}\left(W -  \o +   {i \over 4} \o^{\a\b}  \widehat S_\a\star\widehat S_{\b} {}_{| z = 0}\right) \ = \ \left[W -  \o +   {i \over 4} \o^{\a\b}\widehat S_\a\star\widehat S_{\b} {}_{| z = 0} \ , \ \e_0 \right]_\star.
\ee
We can  therefore decompose $W$ into a Lorentz-connection part $\half \o^{\a\b} \wM^L_{\a\b, | z = 0}$ and Lorentz-tensor parts as follows
\be
W \ = \ \half\o^{\a\b} \left(M_{\a\b}-  {i \over 4}  \{\widehat S_\a,\widehat S_\b \}_{\star\, | z = 0}\right)+ e  \psi_1  + E^{HS} \psi_1  + \O^{HS} \label{decomplorentz}
\ee
where $e,\o,E^{HS}, \O^{HS}$ are independent of $\psi_1$.  They represent the spin-2 vielbein
and Lorentz connection, and their higher-spin counterparts, respectively.

\section{The 3D Sezgin-Sundell solution}\label{Sec:solution}

\subsection{Gauge function method}

The idea behind the gauge function solution method \cite{Vasiliev:1990bu,Bolotin:1999fa,Sezgin:2005pv} is to first solve the zero-curvature and covariant constancy conditions \eq{hseq1}-\eq{hseq3} on ${\cal X}$ via some appropriately chosen gauge function, which absorbs the spacetime dependence of the master fields, and then solve for the twistor-space dependence from the remaining equations. In particular, since $\widehat W$ is flat, it can locally be gauged away by a gauge function $\widehat L=\wL(\ty,\tz;\psi_1,k|x)$. In this ``nothing gauge'' the Vasiliev equations \eq{hseq1}-\eq{hseq5} with the expansion \eq{nplusF} reduce to a kind of deformed oscillator problem for spacetime-independent fields $\widehat \F', \widehat S_\a'$, which is encoded in the ``internal'' equations \eq{internal1}-\eq{internal2}:
\bea
\widehat  W &=&  \wL^{-1}\star d \wL \ ,\\
\widehat \F &=&  \wL^{-1}\star \widehat  \F' \star  \wL, \qquad d  \widehat  \F'  =0 \ ,\\
 \widehat  S_\a &=&  \wL^{-1}\star \widehat  S_\a' \star  \wL, \qquad d  \widehat  S_\a' =0 \ ,\\
\, [ \widehat   S_\a', \widehat   \F']_\star &=&0 \ , \label{internal1}\\
\,[ \widehat  S_\a', \widehat  S_\b' ]_\star &=& -2  i \e_{\a\b} ( 1 + \n \widehat K +  \widehat  \F' \star\widehat K ) \label{internal2}\ .\eea
After solving for $\widehat \F', \widehat S_\a'$  from the last two equations, one can reconstruct the spacetime dependence of all fields by taking the $\star$-products with $\wL$ and its inverse.

\subsection{AdS $\n$-vacua in stereographic coordinates}

The simplest solution is to take the ``nothing gauge'' solution to be given by the deformed oscillator solution of Section \ref{Sec:hsalg}
\bea
\widehat \F' &=& 0\ ,\\
\widehat S_\a' &=& \tilde z_\a \ .\label{nuvacsols}
\eea
If we assume that the gauge function only depends on the $\ty$ oscillators, $\wL = L(\ty;\psi_1,k|x)$, then it follows that $L$ commutes with $\tilde z_\a$ so that we also have $\widehat S_\a = \tilde z_\a$. Thus, from (\ref{Lor}) the Lorentz generators are simply
\be
\wM^L_{\a\b} \ = \ M_{\a\b} \ = \ -{i \over 4} \{ \tilde y_\a, \tilde y_\b \}_\star \ ,
\ee
and the above Ansatz obviously results in a solution with the full $\mso(2,2)$-symmetry (and its higher-spin extension).

Now we turn to the determination of the gauge function $L$ which produces the $AdS$ vacuum without higher spins turned on. $AdS_3$ is the coset $SO(2,2)/SO(2,1)$ and the standard theory of coset manifolds (see e.g. \cite{Gilmore:2008zz}) fixes the gauge function
to be simply the coset representative within $SO(2,2)$. As explained in appendix \ref{App:coset}, in stereographic coordinates this representative is \footnote{This can be
rewritten in the form    used in \cite{Sezgin:2005pv}:
$$L =  \exp_\star \left( - {i \over 4}{ {\rm arctanh} \sqrt{ 1-h \over 1+h }\over  \sqrt{ x^2 } } x^{\a \b} \{ \tilde y_\a, \tilde y_\b \}_\star \psi_1\right)\\$$
where $h = \sqrt{1-x^2}$.}
\bea
L(\tilde y;\psi_1|x) &=& \exp_\star\left( - {i \over 8}{ {\rm arctanh} \sqrt{ x^2 } \over \sqrt{ x^2 }} x^{\a \b} \{ \tilde y_\a, \tilde y_\b \}_\star \psi_1\right)\\
x^{\a \b} &=& \left( \begin{array}{cc} x^0+x^1 & x^2 \\ x^2 &   x^0-x^1 \end{array}\right)\label{Lstereo}
\eea
Working out $L^{-1} \star d L$ as in (\ref{Ldecomp}) indeed gives the vielbein and Lorentz connection for $AdS_3$:
\bea
e_0 &=& {i\over 8} {dx^{\a\b} \over 1-x^2}  \{ \tilde y_\a, \tilde y_\b \}_\star\ ,\\
\o_0 &=& -{i\over 8} {x^{\a \g} dx_\g^{\ \b} \over 1-x^2}  \{ \tilde y_\a, \tilde y_\b \}_\star\ ,\\
E_0^{HS} &=& \O_0^{HS} =0\ ,\\
ds^2 &=& (e_0, e_0) =  {4 dx^a dx_a \over (1- x^2)^2}\ .\label{adsstereo}
\eea

For $\n=0$, we can also write down the Weyl-ordered formula for $L$, using the identity \cite{Kraus:2012uf}
\be
  \exp_\star \left(\half y^T M y \right)= {\rm sech} \sqrt{\det M} \exp  \left( {\tanh \sqrt{\det M} \over \sqrt{\det M}} \half y^T M y\right)\label{gauss}
\ee
leading to
\be\label{Ly}
L(\tilde y;\psi_1|x) \ = \ \sqrt{2 h \over 1 + h}  \exp  \left( - {i \over 2}  {  x^{\a \b}y_\a y_\b  \over 1 + h} \psi_1\right), \qquad h \equiv  \sqrt{1-x^2}.
\ee

\subsection{Sezgin-Sundell-like non-vacuum solutions}\label{Sec:SS}
Having discussed the solutions describing the AdS vacuum, we now want to construct simple solutions where the scalar field is also turned on. As we discussed in section (\ref{Sec:hsalg}), the propagating components of the scalar reside in the sector where $\widehat \F$ is proportional to $\psi_2$. In the gauge function method, the simplest choice is to set the nothing gauge scalar field equal to $\widehat \F' = \m \psi_2$, with $\m$ a real constant, which parallels the construction of the 4D Sezgin-Sundell solutions \cite{Sezgin:2005pv},\cite{Sezgin:2005hf}. To find the corresponding solution for $\widehat S_\a'$,
we observe that all the ingredients in the $\n$-vacuum solutions $\widehat B', \widehat S_\a'$ of the previous subsection  commute with $\psi_2$. Therefore we can generate the desired new nothing gauge solutions by shifting $\n \to \n + \m \psi_2$ in the vacuum solutions (\ref{S0reg}),(\ref{nuvacsols}). This leads to
\bea
% \widehat W' &=& 0\\
\widehat \F' &=& \m \psi_2\label{F'} \\
\widehat S_\a' &=& \r z_\a \left[ 1+ (\n+ \m \psi_2) \int_{-1}^1 ds(1-s) \left( F^-\left((\n+ \m \psi_2)\ln |s| \right) e^{{i \over 2} (1+s) u }  \right. \right. \nonu
&& + \left. \left. F^+\left((\n+ \m \psi_2) \ln |s| \right)e^{{i \over 2} (1-s)u }k \right)\right] \ .\label{primesol}
% y_\a' &=& y_\a + (\n + \m \psi_2 ) w_\a \star K = \tilde y_\a + \m \psi_2  w_\a \star K
\eea
 As explained before, restricting to the bosonic theory we can set $k=\pm1$ in the above expression for $\wS'_\a$. We observe that the corresponding solutions for $\n=0$ then formally coincides with the solution of the internal bosonic equations with $\wF'=\m\psi_2 k$ of the truncated system \eq{hseq1red}-\eq{hseq5red}.

Next we should specify the gauge function $L$. The simplest possibility, which we shall use in this work, is to choose the gauge function to
be identical to the gauge function of the $AdS$ solution,  namely $L(\tilde y;\psi_1|x)$ in (\ref{Lstereo}). Since in the limit $\m\to 0$ the master fields (\ref{primesol}) reduce to their values in
 the $\n$-vacuum of the previous section, our choice for $L$ guarantees that we recover the pure AdS solution when sending $\m \to  0$.
 Nevertheless more general choices reducing to (\ref{Lstereo}) in the limit $\m \to 0$ would also be suitable.

Note that, due to the dependence of the gauge function on $\psi_1$, turning on $\Phi'$ in the physical sector, linear in $\psi_2$,  effectively amounts to breaking the $\mso(2,2)$-symmetry of the $\n$-vacua down to the Lorentz subalgebra, thereby realizing a three-dimensional analogue of the Sezgin-Sundell four-dimensional Ansatz \cite{Sezgin:2005pv}.

\subsection{The scalar profile}\label{secscalar}
The scalar master field in these solutions takes the form
\be
\widehat \F \ = \ \F \ = \  \m L^{-1} \psi_2 \star L.
\ee
and from (\ref{physscalar}) we find for the physical scalar
 \bea
 \phi &=&  \m \,{\rm tr}  (L^{-1}\star L^{-1} )\ .\label{phisolgen}
 \eea
 We should make here an important remark concerning this scalar profile. Since we are using the gauge function method, using for $L$ the same function as for the AdS vacuum,
the master field $W$ is formally unchanged from the vacuum solution, although its decomposition in terms of Lorentz tensors is of course different, see (\ref{decomplorentz}). The equation of motion for the zero form master field $\hat \F \equiv \hat \F (\tilde y; \psi_i |x)$,
being linear, is then precisely the one analyzed in \cite{Ammon:2011ua}, where it was shown that the physical field  $\phi$ satisfies  the free Klein-Gordon equation in global AdS.  While it is known that
the PV theory describes a complex interacting system of a scalar coupled to gravity and higher spins, the gauge function method picks out solutions where the scalar profile
is that of a free scalar in AdS. We will describe a toy example of a scalar-gravity theory with this property in section \ref{sectoy}.

In the simplest case $\n =0$, the gauge function is constructed from undeformed oscillators, $L = L(y;\psi_1|x)$. We can explicitly evaluate $ L^{-1}\star L^{-1}$ using (\ref{gauss}), with the simple answer
\be
 L^{-1}\star L^{-1} =  \sqrt{1-x^2} \exp \left( - {i\over 2}  x^{\a \b} y_\a y_\b \psi_1 \right)
\ee
and we get
\be
\phi = \m  \sqrt{1-x^2} \qquad {\rm for\ } \n=0. \label{phiSS}
\ee
%Converting to Poincar\'e coordinates (see appendix \ref{}) this gives
%\be
%\F_{\n=0} = 2 \m  \sqrt{  z \over y^\m y_\m + (z + 1)^2}.
%\ee
One easily checks that (\ref{phiSS}) satisfies the Klein-Gordon equation with $m^2 = - 3/4$ (i.e., it is an $AdS_3$-massless scalar with $M^2=0$, see footnote 4) with respect to the metric (\ref{adsstereo}), as expected from our remarks above.

For general values of the vacuum parameter $\n$,  the theory is governed by the higher spin algebra  $\mhs[\l ]$  with, projecting onto $k=1$ for definiteness,
\be
\l ={1- \n \over 2} \ .
\ee
We will only consider the range $-1 \leq \n \leq 1, \ 0 \leq \l \leq 1$, for which the PV theory is conjectured to be dual to the 't Hooft limit of the
unitary $W_N$ minimal models \cite{Gaberdiel:2012uj}.
The overall mass of the scalar $\f$ is $m^2 =\l^2 -1$ and the solutions to $m^2 = \D(\D-2)$ are
\be
\D_\pm = 1 \pm \l.
\ee
From (\ref{phisolgen}) we find that the scalar profile in the case of general $\l$ is given by
 \bea
  \f
  &=& \m \, \tr \exp_\star \left( {i \over 4}{ {\rm arctanh} \sqrt{ x^2 } \over \sqrt{ x^2 }} x^{\a \b} \{ \tilde y_\a , \tilde y_\b\}_\star \psi_1\right)\nonu
&=& \m\,  \tr \cosh_\star  \left({i \over 4}{ {\rm arctanh} \sqrt{ x^2 } \over \sqrt{ x^2 }} x^{\a \b}\{ \tilde y_\a , \tilde y_\b\}_\star\right)
\eea
where in the second line we have used that the terms proportional to $\psi_1$ do not contribute to the trace as defined in (\ref{trdef}). The final expression
 is a sum of two traces of group elements in the $SL(2,\RR)$ subgroup of the higher spin symmetry, and in Appendix \ref{apptraces} we provide one way of computing such traces using the fact that $\mhs[\l ]$ can be
 viewed as an $N\to \l$ continuation
 of $\msl(N)$ \cite{Feigin88,Vasiliev:1997dq,Prokushkin:1998bq,Vasiliev:1999ba}.
Using (\ref{charsl22}) with $\sqrt{v^2} = 4 {\rm arctanh} \sqrt{x^2}$ we find
\be
 \f  = \m {\sinh ({2 \l \,{\rm arctanh} \sqrt{x^2} )} \over \l \sinh ({2\, {\rm arctanh} \sqrt{x^2}})}\label{phisollambda}
% &=& {\m ( 1- x^2) \over 4 \l \sqrt{x^2}}\left(  \left( { 1- \sqrt{x^2} \over  1+ \sqrt{x^2}} \right)^{-\l} - \left( { 1- \sqrt{x^2} \over  1+ \sqrt{x^2}} \right)^{\l} \right) .
\ee
For $\l = \half$, this scalar profile reduces to (\ref{phiSS}), and one can also easily check that it satisfies the free Klein-Gordon equation with $m^2 =\l^2 -1$ as expected.

\subsection{A generalization}\label{secgen}
Before continuing our analysis we would like to mention a possible generalization.
%Note, however, that t
The solutions \eq{primesol} can further be dressed with certain Lorentz-invariant projector solutions which are the three-dimensional counterparts of those first found in \cite{Iazeolla:2007wt} and that are here obtained in Appendix \ref{App:defosc}. These solutions, that seem to be gauge-inequivalent to $AdS_3$ (although we shall not give a proof of this statement in this work), are activated by discrete parameters $\th_n=\{0,1\}$, each turning on a contribution from a Fock-space projector $P_n(u)$. Since they survive the limit $\n,\Phi' \rightarrow 0$, they furnish new solutions without matter of the form
\be \wS'_\a \ = \ \rho z_\a \left(1-2\sum_{n=0}^\infty \th_n P_n(u)\right) \ . \label{thetavac} \ee
Splitting $\wS'_\a=\rho(z_\a-2i\wA'_\a)$ into its $AdS_3$ vacuum value, that act as a derivative in $z_\a$ via commutators \eq{yzder}, and a ${\cal Z}$-space connection, the l.h.s. of \eq{internal2} corresponds to the $AdS$-vacuum solution plus a ${\cal Z}$-space field-strength $F'_{\a\b}=2\partial_{[\a} \wA'_{\b]}+[\wA'_\a,\wA'_\b]$. This suggests that these projector solutions represent monodromies of a flat but non-trivial connection on ${\cal Z}$, encoding the interesting possibility that $\wS'_\a$ might carry global degrees of freedom associated to the $z_\a$-oscillators (qualitatively similar to windings). This issue touches upon various other open questions, such as the definition and interpretation of higher-spin observables (see, e.g., \cite{Sezgin:2005pv,Sezgin:2011hq,Vasiliev:2015mka} for the construction of gauge-invariant quantities), as well as on similar solutions of the Chern-Simons theory \cite{Castro:2011iw}. We shall defer a more thorough study of such solutions to a future work, referring the reader to Appendix \ref{App:defosc} for the technical details.

% For example, computing it in the 2-dimensional representation, which corresponds to $\l=2$, we get
% \be
% \f_{\l = 2} = \m { 1 - (x^0)^2 + (x^1)^2 + z^2\over z}.
%\ee

\subsection{Gauge sector for $\n=0$}\label{secgauge}
We shall now proceed to  extract the expressions for  the Lorentz-connection and vielbein components from the master gauge field $W$ in the manner explained in section \ref{seclorentz}. These are a bit more involved, and we will
restrict ourselves to the case $\n =0$ for simplicity.
%We shall now proceed to giving the explicit form of this solution in the undeformed $\n=0$ case, for which the spacetime interpretation of all sectors of the theory is more transparent, leaving for later the generalization of the scalar profile for $\n\neq 0$.
To obtain them we have to decompose $W$ %use the fact that on the one hand, $W$ is formally unchanged from its AdS value, while on the other hand it can be  decomposed
into Lorentz-connection and  Lorentz tensors  according to  (\ref{decomplorentz}). This leads to
\bea
%A \equiv  e  \psi_1 + \o + E^{hs} \psi_1  + \O^{hs} = -  L^{-1} \star d L  +   {i \over 4} \o^{\a\b} \{\widehat S_\a,\widehat S_\b \}_{\star\, | z = \psi_2=0}
W &=& -{i\over 4}  \o_0^{\a\b}y_\a y_\b  -{i\over 4}  e_0^{\a\b}y_\a y_\b \psi_1  \nonumber \\
& =&   -{i\over 4} \o^{\a\b}  \left( y_\a y_\b + {1 \over 2} \{\widehat S_\a,\widehat S_\b \}_{\star\, | z = 0} \right) -{i\over 4}  e^{\a\b}y_\a y_\b \psi_1 + E^{HS} \psi_1  + \O^{HS} \ .
 \label{vbeq}
\eea
The $\wS_\a$-dependent deformation of the Lorentz generators can be evaluated as
\be   {1 \over 2} \{\widehat S_\a,\widehat S_\b \}_{\star} \ = \  {1 \over 2} L^{-1}\star \{\widehat S'_\a,\widehat S'_\b \}_{\star}\star L \ ,\label{LrotS}\ee
inserting the expression \eq{primesol} at $\n=0$
\bea \!\!\!\widehat S_\a' \ = \ \r z_\a \left[ 1+ \m  \int_{-1}^1 ds(1-s) \left( F^-\left(\m \ln |s| \right) e^{{i \over 2} (1+s) u } + F^+\left(\m \ln |s| \right)e^{{i \over 2} (1-s)u }\psi_2k \right)\right] \ .\label{primesol2} \eea
in the r.h.s. of \eq{LrotS} and evaluating the $\star$-products. We collect in Appendix \ref{lemmas} a few relevant steps of the calculation.  Note that the $\psi_2$-dependence of $\wS'_\a$ leads to the appearance of two sectors in $W$: the physical one, proportional to the unit element, and the twisted one, proportional to $\psi_2$, which  we shall now display separately\footnote{Strictly speaking, the notion of ``physical'' and ``twisted'' fields refers to an expansion over the $AdS_3$-background, as recalled in Section \ref{Sec:hsalg}. We keep this distinction here as we look at our exact solution as a deformation of $AdS_3$ turned on by the continuous deformation parameter $\mu$.}.
Defining
\be a^{\a\b}:=\frac{x^{\a\b}}{1+h} \ , \qquad  a^2=\frac{1-h}{1+h} \ ,\ee
one finds, for the physical sector,
\bea
{1 \over 2} \{\widehat S_\a,\widehat S_\b \}_{\star\, | z = \psi_2=0} &=& A \,y_\a y_\b + B \,a_\a^{\ \g}y_\g a_\b^{\ \d}y_\d + C\, a_{(\a}^{\ \ \g}y_{\b)} y_\g \psi_1 \ ,\label{SSphys}\eea
where we have denoted
\bea
A &=& \int_{-1}^1 ds \int_{-1}^1 d\tilde s \,G(s,\tilde s) \left(1 + s \tilde s a^2 \right)^2\ ,\\
B &=& \int_{-1}^1 ds \int_{-1}^1 d\tilde s \,G(s,\tilde s) \left(1 + s \tilde s \right)^2\ ,\\
C &=& 2 \int_{-1}^1 ds \int_{-1}^1 d\tilde s \,G(s,\tilde s) \left(1 + s \tilde s a^2 \right)\left(1 + s \tilde s \right)\ ,\\
G(s,\tilde s) &=& - {\m^2 (1- a^2)\over 4 \left( 1- (s \tilde s)^2 a^2\right)^3} \left[ (1-s)^2 (1-\tilde s)^2 F^+(\m \ln |s|) F^+ (\m \ln |\tilde s|)\right.\nonu
 &&\left.+ (1-s^2) (1-\tilde s^2) F^-(\m \ln |s|) F^- (\m \ln |\tilde s|) \right]\ .
\eea
Thus, the contribution of the non-linear correction to the Lorentz generators only contains bilinear terms in the $y$ oscillators, and therefore it does not give rise to any higher-spin fields. This is similar to the four-dimensional result of \cite{Sezgin:2005pv}.

On the other hand, in the twisted sector, which as previously recalled is peculiar to the three-dimensional theory, we find
\bea
& [{\psi_2  \over 2}& \{\widehat S_\a,\widehat S_\b \}_{\star}]_{\, | z  =  \psi_2=0} \ = \  -k \int_{-1}^1ds\left[F_1\, a_\a^{\ \g}y_\g \,a_\b^{\ \d}y_\d + F_2\,\psi_1 a_{\a\b}\right]\, e^{\frac{i}{1+s^2a^2}\frac{(1+s)^2}{2}a^{\a\b}y_\a y_\b\psi_1}  \nonumber\\
 &&+  k \int_{-1}^1 \,ds\int_{-1}^1 \,d\tilde s\,\left[F_3\, y_\a y_\b+ F_4\,a_\a^{\ \g}y_\g \,a_\b^{\ \d}y_\d \right]\, e^{\frac{i}{1+s^2\tilde s^2a^2}\frac{(1-s\tilde s)^2}{2}a^{\a\b}y_\a y_\b\psi_1} \ ,\label{SStwist}\eea
where
\bea
F_1 &=& \m(1-a^2) \int_{-1}^1 ds \,(1-s^2)F^+(\m\ln|s|) \frac{(1+s)^2}{(1+s^2a^2)^3}\ ,\\
F_2 &=& \m(1-a^2)\int_{-1}^1 ds\,(1-s^2) F^+(\m\ln|s|) \frac{2i}{(1+s^2a^2)^2}\ ,\\
F_3 &=&  \int_{-1}^1 ds \int_{-1}^1 d\tilde s \,H(s,\tilde s)(1-s)(1+\tilde s) \left(1 + s \tilde s a^2 \right)^2\ ,\\
F_4 &=& - \int_{-1}^1 ds \int_{-1}^1 d\tilde s \,H(s,\tilde s)(1+s)(1-\tilde s) \left(1 - s \tilde s \right)^2\ ,\\
H(s,\tilde s) &=&  {\m^2(1- a^2)\over 2 \left( 1+ (s \tilde s)^2 a^2\right)^3}  (1-s) (1-\tilde s) F^+(\m \ln |s|) F^- (\m \ln |\tilde s|)\ .
\eea
Note that the twisted sector contains contributions from all spins. With the goal of attempting a holographic interpretation of this solution, in the following we shall focus on the physical sector, leaving a study of this result for the twisted fields to future work.

Inserting \eq{SSphys} in \eq{vbeq}  and solving for $\o$, $\e$, $E^{HS}$ and $\O^{HS}$ leads to
%
%\bea
%\o^a &=& {1 \over 1 + A + a^2 B} \o_0^a\\
%e^a &=& f_1 dx^a + f_2 x^a d(x^2)\\
%f_2 &=& - {C \over h^2(1+h) ( 1 + A + a^2 B)},\qquad f_1 = {2\over h^2} - 2 x^2 f_2 \\
% E^{HS} &=& \O^{HS} =0.
%\eea
%
%\comment{Replace by formulas below.}
%The solution can be written as
\bea
\o^{ab} &=& f \o_0^{ab}\label{omSS}\\
e^a &=& \h  e_0^a + {(1- \h) x^a \over x^2 (1-x^2)} d(x^2)\label{eSS}\\
E^{HS} &=& \O^{HS} =0 \  \label{3DSS}
%\f &=& \m  \sqrt{1-x^2}\label{phiSS}
\eea
where $e_0$, $\o_0$ denote the global AdS solution
\bea
e_0^a &=& {2 dx^a \over 1- x^2}\\
\o_0^{ab} &=&  -{ 2 ( x^a dx^b- x^b dx^a)  \over 1- x^2}\ ,\label{solads}
\eea
and we defined
\bea
f &=& {1 \over 1 + A + a^2 B}\label{fSS} \ ,\\
\h &=& 1 + {C x^2 f\over 1 + \sqrt{1- x^2}} 	\ .\label{etaSS}
%a^2 &\equiv& {x^2 \over (1 + \sqrt{1-x^2})^2}
\eea
Note that, apart from the necessary adaptations to the three-dimensional formalism, the algebraic structure of this solution is identical to the one of the four-dimensional Sezgin-Sundell solution \cite{Sezgin:2005pv}. The main difference lies in the integral coefficients, which here are slightly more complicated, and is a consequence of the different realization of translations in $AdS_3$ (i.e., of the presence of the Clifford elements $\psi_1$ and $\psi_2$ in $L$ and $\Phi'$, respectively). This accounts for the fact that $f$ and $\h$ are  different functions of $x^2$ from their 4D counterparts.

Finally, note that the solutions above are also solutions of the minimal bosonic PV model discussed in Section \ref{Sec:hsalg}. In order to show this, it is enough to realize that $\tau(L)=L^{-1}$, and, as a consequence, the truncation conditions \eq{tautrunc} are satisfied provided that $\tau(\wB', \wS'_\a) \ = \ (\wB', -i\wS'_\a)$ (the condition on $\wW$ follows automatically from $\wW=L^{-1}\star d L$ and the properties of $L$ under $\tau$). It is immediate to check that this is true for \eq{F'}-\eq{primesol}, as well as for \eq{thetavac}.

\section{Spacetime geometry}\label{secgeom}

In this section we will discuss the spacetime geometry of the our solution (\ref{phiSS},\, \ref{omSS}-\ref{3DSS}).  More precisely, we will discuss geometrical aspects from the  point of view of the spin-2 subalgebra of the higher spin algebra, i.e. treating them as we would solutions of a standard scalar-gravity theory.  A complete discussion would require  a fully higher-spin generalization of differential geometry, which is not available at present.
 Barring this possible caveat, we will argue that
the metric is asymptotically AdS and will see that the connection has torsion, which can be removed by an approprate field redefinition. Finally, we will discuss the interpretation of the solution as a Coleman-De Luccia bubble in AdS.
\subsection{Asymptotically AdS behaviour}
We will presently argue that the solution  (\ref{phiSS},\, \ref{omSS}-\ref{3DSS}) is conformally AdS with the conformal factor approaching one at the boundary. In other words, it is
asymptotically AdS with the same value of the cosmological constant as the global AdS vacuum solution. This suggests that we should be
able to holographically interpret the solution in (a possibly  relevant or marginal deformation of) the CFT whose vacuum state is represented in the bulk by the
global $AdS$ solution (\ref{solads}). We will comment on the holographic interpretation of the solution in section \ref{sechol}.

The conformally AdS character of the solution is manifest after rewriting  the vielbein (\ref{eSS}) as
\be
e^a = {2 \O\, d( g_1 x^a) \over 1 - g_1^2 x^2}.
\ee
where
\bea
g_1 &=& \exp {\int_1^{x^2} {1-\h (t) \over 2 \h(t)  t} dt} \ ,\label{g1SS}\\
\O &=& {1 - g_1^2 x^2 \over g_1( 1- x^2)} \h \ .
\eea
In solving for $g_1$, we have chosen  the boundary condition that $\lim_{x^2 \to 1} g_1=1$.

The conformal boundary is still at  $x^2 \to 1$ since, as we shall presently see, the  conformal factor $\O$ is regular there. The limiting value
$\lim_{x^2 \to  1} \O $  determines the  asymptotic value of the AdS radius. To determine it we need the asymptotic behaviour of
the quantities defined above. Defining  $h \equiv \sqrt{1-x^2}$ we find, working to order $\m^2$ for simplicity, the small-$h$ expansions
\bea
A &\sim &- {\m^2 \over 8} + {\p^2 \m^2\over 128} h+ \calo (h^2, \m^4)\label{nb1}\\
B &\sim & - {\m^2 \over 8} + {(\p^2-32) \m^2\over 128} h + \calo (h^2, \m^4)\\
C &\sim & - {\m^2 \over 4} + {(\p^2-16) \m^2\over 64} h + \calo (h^2, \m^4)\label{nb3}
\eea

Substituting in (\ref{fSS}, \ref{etaSS}, \ref{g1SS})  leads to
\be
\O = 1 + {\p^2 \m^2 \over 64} h +  \calo (h^2, \m^4)
\ee
Hence at least to the first nontrivial order in $\m$ the asymptotic value of the  cosmological constant is the same as for the vacuum solution\footnote{The same conclusion holds for the 4D Sezgin-Sundell solution, contrary to what was claimed in eq. (4.68) of \cite{Sezgin:2005pv} which we believe to be erroneous.
%. We believe that this is an error due to neglecting  the $\calo (h^2)$ contribution to $g_1$ which however contributes to the leading behaviour of $\O$.
}.

\subsection{Torsion and field redefinitions}
Before studying the geometry of our solutions, we would like to discuss a subtlety related to field redefinitions. Since we do not yet have an action at our disposal for our theory\footnote{See however \cite{Vasiliev:2015mka, Boulanger:2011dd} for interesting proposals for constructing an action for the 3D and 4D Vasiliev system.},
we also do not know which is the preferred set of  fields in terms of which the action takes a canonical form. For our class of solutions, any redefinition of the fields $\f, e^a, \o^{ab}$ which preserves their transformation under reparametrizations and local Lorentz transformations is just as good.  %In this sense, it was preliminary to

In the current field redefinition frame, the connection defined by (\ref{omSS}) has torsion. Indeed, one finds it to be of the form
\be
T^a \equiv de^a  +\o^{a}_{\ b} e^b = \left( (\ln \h)' + l \right) d(x^2 ) \wedge e^a\label{torsion}
\ee
where a prime denotes differentiation with respect to $x^2$, and $l$ is the function
\be
l(x^2) = {\h-1 + x^2 (1 + \h - 2 f)\over 2 x^2 (1-x^2)\h }.
\ee
Here we see another difference with the 4D Sezgin-Sundell solution where the combination $l$ actually vanishes, whereas in the case at hand   it does not.
In fact, the function $l$ is singular near the boundary, where it has the expansion
\be
 l(x^2) \sim \frac{\pi ^2 \mu ^2}{32 h}-\frac{\mu ^2}{2 h^2} +\calo(h^0, \m^2).\label{lSSblowsup}
 \ee

For comparison with the standard framework for holographic duality it would certainly be preferable to work in a field redefinition frame where the torsion vanishes.
There is actually an infinitude of field redefinitions which accomplish this. One such redefinition, which was chosen for the 4D Sezgin-Sundell solution in \cite{Sezgin:2005hf}, is  to Weyl-rescale the vielbein while keeping the connection and the scalar field unchanged.
However, it is not hard to see that in the present case, due to the singular behaviour of $l$ in (\ref{lSSblowsup}), the required conformal factor blows up near the boundary, destroying the nice asymptotically AdS behaviour of the metric found in the previous section. Therefore we will propose to instead  redefine the Lorentz connection, leaving the vielbein and scalar unchanged,
 in the following way. The torsion (\ref{torsion}) can be written as
 \be
 T^a = d H \wedge e^a
 \ee
 with $H$ a certain function of $x^2$. Because the scalar profile is monotonic in $x^2$, we can also view $H$ as a function of $\f$, and hence $T^a$ as constructed out of
 $\f$ and $e^a$.
 Similarly, the contorsion tensor
 \be
 K_{\m[\n\r]} = - \half \left( T_{[\m\n]\r}- T_{[\n\r]\m}+ T_{[\r\m]\n}\right)
 \ee
 can be viewed as constructed out of $\f$ and $e^a$.
 We then propose to make the field redefinition
 \be
\o^{ab} \to \tilde \o^{ab} (\o, e, \f) = \o^{ab} - K^{ab} ( \f, e),
 \ee
 while leaving $\f$ and $e^a$ unchanged.
 When evaluated on our particular solution, $\tilde \o$ becomes by construction the Levi-Civita connection for the vielbein $e^a$. Hence in our proposed field redefinition frame we can keep working with
 the vielbein (\ref{eSS}), while spacetime curvature is measured by the standard Levi-Civita connection.
\subsection{Interpretation as a Coleman-De Luccia bounce-bubble-crunch}
Let us now discuss the geometry of our solution in more detail. We start by discussing the solution in Euclidean signature\footnote{It was shown in \cite{Iazeolla:2007wt} that, at least in four dimensions, the analytic continuation
between Euclidean and Lorentzian signatures is compatible with the Vasiliev equations.}. We introduce spherical coordinates
\be
x^a = \tanh {\r  \over 2} n^a\label{spherical}
\ee
with $n^a n_a=1$. The solution then takes the form
\bea
ds^2 &=& d\r ^2 + \h(\r)^2 \sinh^2 \r   ( d\theta^2 + \cos^2 \theta d\vf^2)\\
\f &=& \m \, {\rm sech} {\r \over 2}\label{Phieucl}
\eea
and is asymptotically Euclidean AdS in a spherical slicing.
Furthermore the solution is regular at the center $\r =0$
since one can check that $\h(0)=1, \f'(0)=0$.

The solution has the characteristic form of an $O(3)$-invariant bounce, describing  the materialization of a Coleman-De Luccia bubble within a metastable AdS vacuum, as discussed in their classic paper \cite{Coleman:1980aw}.  From the scalar profile (\ref{Phieucl}) we see
 that  our solution does not describe a thin-walled bubble, rather the scalar field is only excited  in a small region near the origin $\r =0$.
The existence of the bounce solution indicates that it falls within  boundary conditions which render the AdS vacuum unstable. Examples of such boundary conditions in AdS gravity
have been studied extensively in the literature starting from \cite{Hertog:2004rz} (see \cite{Hertog:2005hu}-\cite{Barbon:2011ta} for a partial list of further references), and our solutions appear to be higher spin gravity analogues of the  bounce solutions appearing in those works. We will have more to say on this in the next section.

The solution in Minkowski signature, in the patch where $x^2>0$, can be written in coordinates analogous to (\ref{spherical}) as
 \bea
ds^2 &=& d\r ^2 + \h(\r)^2 \sinh^2 \r   ( - d\t^2 + \cosh^2 \t d\vf^2)\\
\f &=& \m \,{\rm sech} {\r \over 2}\label{bubblepatch}
\eea
It has manifest $O(2,1)$ symmetry and  is asymptotically Lorentzian AdS in a de Sitter slicing. It can be matched to the Euclidean solution
at $\theta = \t=0$ and then describes the growth of the bubble after it materializes. Alternatively, we can forget about the bounce and extend the solution (\ref{bubblepatch}) to
negative values of $\t$ .

The above coordinates break down at $\r =0$; the solution can be continued to the patch $x^2<0$  inside the bubble by choosing the coordinates
\be
x^a =  \tan {T \over 2} m^a
\ee
where $m^a m_a = -1$. This leads to
 \bea
ds^2 &=& - dT^2 + \h(T)^2 \sin^2 T  ( d\b^2 + \sinh^2 \b d\vf^2)\\
\f &=& { \m \over \cos {T \over 2}}
\eea
This is an open FRW universe whose constant time slices are copies of the hyperbolic plane $H_2$. The spacetime undergoes a big crunch
at $T = \p$\footnote{This result is to be compared with the conclusions of \cite{Sezgin:2005hf} where, in four dimensions and in a different torsion-free frame, the analogue of this solution was found to be singular in the Type B model and singularity-free in the Type A model.}. When the scalar is turned off, $\m=0$, this is just a coordinate artifact, but for $\m \neq 0$ it is a genuine
singularity as can be seen from the fact that $\f$ blows up. We should however keep in mind that we used the term `singularity' here in the standard
differential geometry sense, in that there exists a diffeomorphism invariant quantity which diverges. This does not necessarily imply that the solution is singular in the sense of higher spin geometry, i.e. that there exists  a  higher spin invariant quantity which diverges on the solution. We defer the study of this issue to future work.

\section{Towards a holographic interpretation}\label{sechol}
In this section we initiate the the study of our solutions from a holographic viewpoint. We start by displaying their near-boundary behaviour
and comparing it with that found in standard scalar-gravity theories, addressing another subtlety related to field redefinitions. Despite the
fact that the precise holographic dictionary for the full PV system is currently unknown, we will go on to propose a holographic picture for
our solutions relying on some mild assumptions on the AdS/CFT dictionary. This proposed interpretation is analogous to the holographic
picture which has emerged for big crunch solutions in other holographic systems  \cite{Hertog:2004rz}-\cite{Barbon:2011ta}. Finally, we will study in more detail the limit $\l \to 0$ of our
solutions, where the new phenomenon of a running coupling in the dual theory emerges.

\subsection{Comparison with 2-derivative scalar-gravity theories}\label{sectoy}

In this section we will compare the near-boundary behaviour of our solutions (\ref{eSS},\ref{phiSS}) with that of standard two-derivative scalar-gravity theories.
Using the near-boundary expansions (\ref{nb1}-\ref{nb3}) and performing a shift of the radial coordinate
\be
 \r = \tilde \r  + {\m^2 \over 4} + \calo (\m^4 )
\ee
we  find the following near-boundary behaviour of our solutions (\ref{bubblepatch}):
\bea
ds^2 &=& d\tilde \r  ^2 + G^2(\tilde \r  )  ds^2_{dS_2}\label{rhocoord}\\
G^2(\tilde \r  ) &\sim&  {e^{2 \tilde \r   }\over 4}  + \frac{ \pi ^2 \mu ^2}{64} e^{3 \tilde \r \over 2 } + \calo \left( e^{\tilde \r   }, \m^4 \right)\label{metras}\\
\f &\sim& 2 \mu\left(1 -\frac{\mu ^2}{8}\right){e^{-{\tilde \r   \over 2}  }} - 2 \m \left(1-\frac{3 \mu ^2}{8}\right) e^{-{3 \tilde \r   \over 2}} +  \calo \left( (e^{-{5 \tilde \r  \over 4} }, \m^4 \right)\label{Phias}
\eea
The shift of $\r$ was performed so that the leading $e^{2 \tilde \rho }$ factor multiplies the $dS_2$  metric with curvature radius $\half$, independent of $\m$.

The scalar profile has the standard falloff of a massive scalar field
\be
\f \sim \b e^{- \D_- \tilde \r  } + \a  e^{- \D_+ \tilde \r  } \label{Phisasgen}
\ee
where $\D_\pm$ are the roots of the equation $m^2= \D (\D -2)$ and $m^2  = - 3/4$.
This behaviour is  guaranteed for solutions constructed using our  gauge function, as we saw in section \ref{secscalar}.

On the other hand, the subleading behaviour of the metric in (\ref{metras}) grows faster near the boundary than one might naively expect, and
 we will devote the rest of this section to understanding it. Let us compare the asymptotic  behaviour   (\ref{metras}) to the one encountered in scalar-gravity theories, which we will first review. Consider a canonically normalized scalar $\chi$ coupled to Einstein gravity through a two-derivative action
 \bea
 S &=& \int d^3 x\sqrt{-g} \left( R + 2 - \e \left({1 \over 2} \pa_\m \chi \pa^\m \chi + V(\chi) \right) \right) \nonu
V( \chi ) &=& {m^2\over 2} \chi^ 2 + \calo (\chi^3)\label{canaction}
\eea
 where for later convenience we have introduced a sign factor $\e = \pm 1$; positive kinetic energy requires $\e = 1$.
 We are interested in  $O(2,1)$-invariant solutions; it is convenient to parametrize them in terms of a new radial coordinate $\tilde r$ in the following way:
 \bea
 ds^2 &=& {a(\tilde r)^2 d\tilde r^2 } + \tilde r^2 ds^2_{dS_2}\ ,\\
 \chi &=& \chi(\tilde r) \ .
 \eea
 The Einstein equation determines the function $a$ to be
 \be
 a(\tilde r)^2 \ = \ {\e \tilde r^2 (\chi')^2 -4\over 2 \e \tilde r^2 V(\chi) - 4 (\tilde r^2 +1)}\ .
 \ee
Inserting the asymptotic expansion of the scalar
\be
\chi \sim \tilde \b \tilde r^{-\D_-} + \calo (\tilde r^{\D_--2})
\ee
we find, for $\D_- <1$, the metric behaviour
 \be
a(\tilde r)^2 \sim \tilde r^{-2}  - {\e \tilde \b^2 \D_- \over 4} \tilde r^{-2(1+\D_-)} + \calo (\tilde r^{max(-4,1- 4 \D_-)})\label{asa}
 \ee
For $\e=1$, this is precisely the asymptotic behaviour of a solution satisfying the  conformally invariant boundary conditions introduced by Hertog and Maeda \cite{Hertog:2004dr}.
Translating back to the radial coordinate $\tilde \r  $ in terms of which the metric takes the form (\ref{rhocoord}), (\ref{asa}) corresponds to
\be
G^2(\tilde \r  ) \sim  {e^{2 \tilde \r   } \over 4} - {\e \tilde \b^2 \D_- \over 2^{2(1- \D_-)}} e^{2(1- \D_-) \tilde \r   } + \calo \left( (e^{\tilde \r   })^{max(0,2- 4 \D_-)}\right).
\ee

This leads to an apparent puzzle: for a scalar of $m^2 = -3/4, \D_- = 1/2$, the subleading term goes as $e^{\tilde \r  }$ rather than the $e^{3 \tilde \r   \over 2 }$ behaviour found in our Prokushkin-Vasiliev solutions
(\ref{metras}). The resolution  comes from taking into account field redefinitions: indeed,  it is not consistent to model $\f$ as a scalar which has a two-derivative action {\em and} is canonically normalized\footnote{Of course, if the action for $\phi$ would contain higher derivatives, as would be the case in the actual Vasiliev theory, the two-derivative part could still be canonically normalized.}.
To show this, we will now construct
 a two-derivative scalar-gravity toy model which allows precisely (\ref{eSS},\,\ref{phiSS}) as a solution. A similar model was constructed for the 4D Sezgin-Sundell solution in \cite{Sezgin:2005hf}.
We start from a general two-derivative action
 \be
 S = \int d^3 x\sqrt{-g} \left( R + 2 - {1 \over 2} K(\f) \pa_\m \f \pa^\m \f - V(\f)  \right).\label{toyaction}
 \ee
Requiring that (\ref{eSS},\ref{phiSS}) solves the equations of motion fixes $K$ and $V$ to be
\bea
K &=& \frac{\sinh ^2  \rho \, \text{csch}^6\left(\frac{\r  }{2}\right) \left(\eta^2-1 + \sinh^2 \r  \left( (\pa_{\r } \h)^2 - \h \pa_{\r }^2 \h \right)\right)}{2 \mu ^2 \eta^2}\\
V &=&    -\frac{\eta  \pa_{\r }^2 \eta + (\pa_{\r } \eta)^2+4   \coth \r  \, \eta \pa_{\r } \eta +\text{csch}^2 \r  \left(\eta^2-1\right)
    }{\eta^2}
\eea
We are now to rewrite the right-hand sides expressing $\r $ in terms of $\f$ by inverting (\ref{bubblepatch}). In the near-boundary regime, $\f$ is small and we can use the asymptotics (\ref{nb1}-\ref{nb3}) to find
\bea
K &\sim & -{ \p^2 \m \left(1 + {\m^2 \over 4}\right) \over 32 \f} + \calo ( \f^0 , \m^3)\\
V &\sim &{7 \p^2 \m \over 256} \left(1 + {\m^2 \over 4}\right) \f + \calo ( \f^2 , \m^3).
\eea
From the expression for $K$ we see that $\f$ is not canonically normalized. After making the field redefinition
\be
\f = {8  \left(1 - {\m^2 \over 4}\right) \over \p^2 \m}\chi^2 + \calo (\chi^3, \m)
\ee
the action becomes of the form (\ref{canaction}), but now with $m^2 = -7/16$, or $\D_- = 1/4$. This explains  $e^{3 \tilde \r   \over 2 }$ behaviour in our metric (\ref{metras}). Furthermore, since
$K$ is negative, $\e=-1$ in (\ref{canaction}) and the field $\chi$ in our two-derivative theory is a wrong-sign scalar (hence the nonstandard sign in front of the $e^{3 \tilde \r   \over 2 }$ term in (\ref{metras})
as compared to \cite{Hertog:2004dr}). Hence the two-derivative theory, which we cooked up to reproduce a particular PV solution, should not be taken  seriously as an approximation of PV theory: in the two-derivative model, the vacuum is perturbatively unstable, in contrast to the   full PV system which is believed to be  dual to a unitary CFT. This is perhaps not surprising
in light of the fact that the PV system is known to describe higher derivative interactions with arbitrarily high number of derivatives.
%Therefore a detailed holographic understanding of our solutions will likely require knowledge of holographic renormalization for the full Prokushkin-Vasiliev theory.

  \subsection{Dual picture as a CFT runaway}
  In the gauge-gravity correspondence  one typically uses the process of holographic
renormalization \cite{de Haro:2000xn} to extract finite CFT quantities, such as VEVs of various operators, from classical bulk solutions. Since this procedure relies heavily on knowledge of the action of the bulk theory, it has so far been performed only for the subsector of the Prokushkin-Vasiliev where the scalar field is turned off, which can be described by a Chern-Simons action, but not yet for the full theory.
 We have for example at present  no way to read off the VEV of the dual stress tensor for our solutions, which on physical grounds we expect to be zero since they represent what global AdS decays into.
Since the precise holographic dictionary depends sensitively on the off-shell action, it would also not be prudent to try to interpret our solutions using the  holographic renormalization of the scalar-gravity action (\ref{toyaction}), which as we already argued is not a very reliable guide to the full PV system.

However, we have already observed
that, to first order in $\m$, the background remains global AdS while the scalar profile solves the free Klein-Gordon equation in AdS.
We will use this to propose a dual picture for our solutions, under the assumption that the standard AdS-CFT dictionary for free
scalar fields remains valid.
Although this interpretation strictly speaking holds only in the small-$\m$ limit, we expect    the qualitative picture to extend
 to finite $\m$.

 Returning to the case of general $\l$, the scalar profile (\ref{phisollambda}) reads, in terms of the radial coordinate defined in (\ref{spherical}):
\be
 \f  = {\m \sinh \l \r  \over \l \sinh \r }\label{Phinu}.
 \ee
For $\l >0$, the near-boundary behaviour is\footnote{The $\l \to 0$ limit is special and will be discussed in the next section.}
\be
\f \sim  \b e^{- \D_- \r } + \a e^{- \D_+ \r }
\ee
with
\be
\b = - \a = {\m \over \l}. \label{alphabetabubble}
\ee
Since $\f$ satisfies the free Klein-Gordon equation it seems reasonable to assume that the standard AdS/CFT dictionary for a free scalar, and in particular  Witten's prescription for boundary conditions describing multi-trace deformations \cite{Witten:2001ua}, is applicable to our solution in the small-$\m$ limit.
Namely,  imposing a boundary condition on the scalar where the functional form of $\a$ is fixed in terms of $\b$,
\be
\a = \a (\b)\ , \label{mulitrbc}
\ee
corresponds in the dual theory to adding a deformation
\be
\D S = - N \int d^2 x \,W( \calo_- )\label{defS}
\ee
where $\calo_-$ is an operator of dimension $\D_-$ and the function $W (\b )$ satisfies
\be
{\pa W \over \pa \b} = \a (\b ).
\ee
The profile $\b$ is proportional the VEV $\langle \calo_- \rangle$ in the deformed theory, for example taking $\a (\b) =0$ corresponds to the `alternate'  quantization of the scalar field.

We can of course interpret our solutions within an infinitude of different boundary conditions of the form (\ref{mulitrbc}), leading to different dual interpretations of the same bulk solution. In general, such multitrace boundary conditions and their corresponding CFT deformations break conformal invariance unless $W( \calo_- )$ happens to be marginal. We will here choose our boundary conditions precisely such that $W( \calo_- )$
is  marginal, as in the majority\footnote{An alternative interpretation in a field theory with a relevant deformation is discussed in the Appendix of \cite{Maldacena:2010un}.} of works discussing similar solutions.
That is, for $\l > 0$ we will impose the boundary condition \cite{Hertog:2004dr}
\be
\a (\b ) = f \b^{\D_+ \over \D_-}\label{confbc}
\ee
corresponding to deforming the theory with the  marginal\footnote{To be more precise, the operator is classically marginal, but quantum corrections at large $N$ could in principle modify this. We see no evidence for this in the bulk theory except in the $\l=0$ case to be discussed in the next section.}   operator
\be
W(\calo_- ) = {f \D_- \over 2} \calo^{2\over \D_-} \label{CFTdef}
\ee
The boundary condition (\ref{confbc}) is manifestly scale invariant since the parameter $f$ which defines it does not change under the scale transformation $\r\to \r + \L, \b \to e^{\D_- \L}\b$.
In our solutions, we see from (\ref{alphabetabubble}) that $f$  takes the value
\be
f = - \left({\m \over \l}\right)^{2\l \over \l-1}.\label{fval}
\ee
It will be important later on that $f$ is negative.

In the coordinates (\ref{rhocoord}), describing a patch of the asymptotically AdS region (\ref{bubblepatch}), the VEV $\b$ is constant and the CFT description seems smooth, despite the fact that the
bulk solution has a big crunch singularity.
This apparent regularity is  however misleading since, as was pointed out in \cite{Hertog:2004rz}, this patch does not include the event of the  bubble hitting the AdS boundary at $\t = \infty$. To continue beyond
this patch, it is convenient  to go to global coordinates $(t, r, \vf)$ in terms of which  the AdS metric is
\be
ds^2 =- \cosh^2 r\, dt^2 + dr^2 + \sinh^2 r \,d\vf^2 \ .
\ee
From (\ref{global}) we find that our earlier radial coordinate $\r $ corresponds to
\be
\sinh^2 \r  = \sinh^2 r (1- \sin^2 t  \coth^2 r ).\label{toglobal}
\ee
Substituting this into the scalar profile (\ref{bubblepatch}) we find that $\f$ still obeys   the boundary conditions (\ref{confbc}) for the same value (\ref{fval}) of $f$, but now
the VEV $\a$ is  time-dependent while still spatially homogeneous:
\be
\b = {\m \over \l (\cos t)^{\D_-} }.\label{betacrunch}
\ee
The materialization of the bubble happens at global time $t=0$, while the bubble reaches the boundary in finite time  $t = {\p \over 2}$ . At this time, the VEV $\b$
blows up, signalling that the deformation (\ref{CFTdef}) has rendered the  field theory singular.

To understand the behaviour of the VEV (\ref{betacrunch}) from the point of view of the dual theory, we consider the effective action
$\G [\s ]$ for $\s \equiv \langle \calo_- \rangle$ in the theory deformed by (\ref{confbc}), which is defined as the Legendre transform of the generating function of $\calo_-$ correlators.
As is the case for the similar scalar-gravity solutions \cite{Hertog:2004rz}-\cite{Barbon:2011ta}, one can obtain a (surprisingly good) qualitative picture  from the leading terms in an expansion of $\G[\s ]$ in derivatives as in the setup of \cite{Coleman:1973jx}. To second order in derivatives, $\G [\s ]$ is of the form \cite{Elitzur:2005kz}
  \be
{1 \over N} \G [\s ]  = -   \int d^2 x \left( {f \D_- \over 2} \s^{2\over \D_-} +  c { \pa_\m \s \pa^\m \s \over 2 \s^2 }    +\ldots \right).\label{effect}
 \ee
% where the additive constant has been chosen for
Since $\calo_-$ is marginal, the effective action should be conformally invariant, which fixes the form of the second term, with $c$  an unknown constant.
Note that upon redefining \be \s = e^{\sqrt{c} \psi}\ee one obtains the classical Liouville action:
 \be
 {1 \over N} \G [\s ]  = -   \int d^2 x \left( \pa_\m \psi \pa^\m \psi + {f \D_- \over 2} e^{\g \psi} \right)\label{effectliouv}
 \ee
 where
 \be
 \g = {2 \sqrt{c} \over \D_-}.
 \ee
 It will be useful to recall the expression for the stress tensor for Liouville theory defined on the cylinder (see e.g. \cite{Seiberg:1990eb}):
 \be
 T_{++} = \half (\pa_+ \psi)^2 - {1 \over \g} \pa_+^2 \psi + {1 \over 2 \g^2}\label{Tliouv}
 \ee
 and similarly for $T_{--}$, where $x_\pm = \vf \pm t$. The last term, which would be absent on the plane, comes from the Schwarzian derivative
 of the  conformal transformation from the plane to the cylinder.

 When $f$ is negative, the deformed theory is unstable since the potential term in (\ref{effect}) and (\ref{effectliouv}) is unbounded below.
 One easily checks that the effective theory then allows for solutions
 \be
 \s =\left(   {(-f) \cos^2 t  \over c \D_- } \right)^{-{\D_-\over 2}}\label{sigmasol}.
 \ee
 This is precisely of the form of the bulk profile $\b$ in (\ref{betacrunch}) (and coincides with  it for $c= {\m^2 \over \D_- \l^2}$), so it seems plausible that the behaviour of the VEV $\s$ in the theory deformed by (\ref{CFTdef}) with a negative coefficient gives an (at least qualitatively correct) dual description of our bulk solutions.
 In the solution (\ref{sigmasol}),  $\s$ emerges at time $t=0$  at rest at position  $\s_0 = \left(- {c \D_- \over f} \right)^{\D_- \over 2}$ on the slope of the negative potential in (\ref{effect}). The field then rolls down the hill, and since the potential is unbounded below,  $\s$ blows up in finite time $t = \p/2$ when the bubble hits the boundary.
 The stress tensor (\ref{Tliouv}) of these solutions vanishes, as one would expect for a solution that  the vacuum tunnels into.
 Note that our expansion in derivatives (\ref{effect}) is reliable only for $|\pa \s|\ll \s^{1 + 1/\D_-}$, which is the case close to the crunch time $t=\p/2$, but would be ill-suited to study small fluctuations around  the vacuum $\s=0$.

 It would be interesting to apply methods used in the literature to further investigate the singular behaviour of our dual field theory. An interesting question is whether, along the lines of \cite{Hertog:2005hu,Turok:2007ry,Craps:2007ch},  a quantum mechanical treatment favors a 'bounce'  through the singularity into a big bang phase of the evolution. Another approach, also discussed in \cite{Hertog:2005hu}, would be to 'repair' the unboundedness of the potential (\ref{effect}) by adding positive terms which dominate at large $\s$.

 Let us also briefly comment on what our solutions look like in Poincar\'e coordinates $(z,y^0, y^1)$, in terms of which the AdS metric is
  \be
  ds^2 ={ dz^2 + dy^\m dy_\m\over z^2}.
  \ee
 The dual theory is now defined on the plane instead of the cylinder.
 Using
  the transformation (see (\ref{Poincare}))
  \be
  x^2 = {y^\m y_\m + (z-1)^2 \over y^\m y_\m + (z+1)^2 }
  \ee
 we find that the scalar profile once again  fits  in the boundary conditions (\ref{confbc},\ref{fval}), where $\b$ is now given by
 \be
 \b = {\m \over \l} {1 \over (1 +y^\m y_\m )^{\D_-}}.
 \ee
 The corresponding  solution of the effective theory (\ref{effect}) is
  \be
 \s = e^{\sqrt{c} \psi}=\left(   {(-f) (1 +y^\m y_\m  )^2 \over c \D_- } \right)^{-{\D_- \over 2}} \label{sigmasolpoinc}.
 \ee
 This profile is reminiscent of that of the Fubini instantons \cite{Fubini:1976jm} which exist in classically conformally invariant scalar theories in  $d>2$.
 In Appendix \ref{appfubini} we show that the solutions
 (\ref{sigmasolpoinc}) arise as a particular $d\to 2$ scaling limit of the Fubini instantons.

 We end this section with the remark that an analogous holographic interpretation was proposed for the Sezgin-Sundell solution
 in 4D Vasiliev theory, namely as a runaway in a CFT destabilized by a marginal triple-trace deformation \cite{Sezgin:2005hf}. Similar  arguments using the effective action $\G[\s]$ can be applied there as well.

\subsection{The limit $\l \to 0 $ and interacting fermion models}

%We end this note with some preliminary comments on the interesting $\l \to 0$ limit, which we hope to study in more detail in the future.

We now comment on the limit $\l \to 0$ which is particularly interesting since new phenomena appear. On the bulk side, the mass of the scalar then saturates the Breitenlohner-Freedman bound,
$m^2 = -1$, where the scaling dimensions become equal, $\D_+ = \D_- = 1$. For this value of the mass, the profile of  a free scalar  has a logarithmic term in the near-boundary expansion:
 \be
 \f =  \a  \ln (mz)z + \b z
 \ee
 where $z = e^{-  r}$ and we have introduced a scale $m$ to define the logarithm. On the dual side, this corresponds to the fact that the double-trace deformation with the operator dual to $\f$ is only classically marginal, while quantum mechanically it has a running coupling. At large $N$, the corresponding beta function receives only a 1-loop  contribution \cite{Witten:2001ua}.

 The limit  $\l \to 0$ of our scalar profile (\ref{Phinu}) reads
 \be
 \f = {\m  \r \over \sinh \r}
 \ee
 and in global coordinates (\ref{toglobal}) leads
 to
 \bea
 \a &=& -{2\m \over \cos t}\ , \\
 \b &=&-\a \ln ( m \cos t)\ . \label{ablambda0}
 \eea
We will choose to interpret the solution within the boundary  conditions\footnote{An alternative way to interpret the solutions is in the conformally invariant boundary condition of Hertog and Maeda \cite{Hertog:2004dr},
\be
\b (\a ) = \a( \tilde f + \log {\a \over m})\label{bcBF}
\ee
with $\tilde f$ a constant.
The field theory interpretation of these boundary conditions is not entirely clear to us, however.}
\be
\a = f \b
\ee
corresponding to a CFT deformation (\ref{defS}) with a double-trace operator
\be
%{1\over N} \D S = -  \half \int d^2x {f } \calo^2.
W(\calo) = {f \over 2} \calo^2. \label{dtracdef}
\ee
From (\ref{ablambda0}) we see that, in contrast with the $\l>0$ case, $f$  is now a scale- and time-dependent source:
\be
f(m,t) = - {1\over \log (m \cos t)}.\label{flambda0}
\ee
For a fixed scale $m$,  $f$ is small and positive sufficiently close to  the `crunch time'  $t = \p/2$. It is interesting to note that the scale dependence of $f$ is like   in an infrared free theory with  a
positive 1-loop beta function
\be
m \pa_m f(m,t) = f(m,t)^2.\label{betaf}
\ee
and that approaching the crunch time $t \to \p/2$ is equivalent to approaching the infrared regime.

We would now like to explain the bulk relation (\ref{betaf}) and the time-dependent profiles $\a$ and $\beta$ from the point of view of the dual theory.
We can be quite explicit in this case, using the fact that   the undeformed  dual CFT at $\l=0$ is (the singlet sector of) a theory with $N$ free fermions $\Psi^a,\ a = 1,\ldots , N$ \cite{Gaberdiel:2011wb,Gaberdiel:2013jpa}. The operator $\calo$ dual to
$\f$ corresponds to the single-trace singlet operator
\be
\calo = {\sqrt{\p} \bar \Psi^a\Psi_a \over N}.
\ee
where the normalization is chosen so that the beta-function for the  double-trace coupling will turn out to have  coefficient one as in (\ref{betaf}).   Including a double-trace deformation (\ref{dtracdef}) the action reads\footnote{Our 2D spinor conventions are as follows: the $\g$-matrices satisfy $\{ \g_\m, \g_\n \} = 2 \h_{\m\n}$ with $\h = {\rm diag} (-1,1)$, and $\bar \Psi$ stands for $- i \Psi^\dagger \g^0$.}
\be
S = \int d^2 x\left[   \bar \Psi^a \g^\m \pa_m \Psi_a - {\p f \over 2 N} (\bar \Psi^a\Psi_a )^2\right].\label{classferm}
\ee
As usual it's convenient to perform a Hubbard-Stratonovich transformation to the equivalent action
\be
S =  \int d^2 x\left[   \bar \Psi^a ( \g^\m \pa_m  - \s)\Psi_a + {N \over 2 \p f } \s^2\right]\label{hubbstrat}
\ee
The equation for $\s$ is $\s =  \sqrt{\p } f \calo$ so that on the bulk side  we should identify $\s$ with
\be
\s \sim  \a.\label{sigmaalpha}
\ee
If $f$ were small and negative, the leading part of the potential in (\ref{hubbstrat}) would have the `right' sign, and   the action (\ref{classferm}) would  describe the Gross-Neveu model \cite{Gross:1974jv}  which  is asymptotically free. In view of (\ref{flambda0}) we will here consider positive $f$, for which the leading part of the  potential is unbounded below;  in this case the theory is free in the infrared, as we will presently review.

As in the previous subsection, we would like to  compute the effective action for $\s$ in an expansion in derivatives.
Integrating over the $\Psi^a$ fields we obtain the effective action
\be
\G[\s] =  N \int d^2 x\left[  {1 \over 2 \p f } \s^2 - i \tr \ln ( \g^\m \pa_\m - \s) \right]
\ee
For the term without derivatives, the effective potential,  we find in the standard manner \cite{Gross:1974jv}\footnote{We have here neglected finite-size effects coming from the fact that the boundary theory is defined on a cylinder of radius $R_{S^1}= \half$, which is justified since
we are interested in the large field regime $\s R_{S^1} \gg 1$.}
\be
{1\over N} V[\s] = -{\s^2 \over 2 \p f} + {\s^2\over 4\p}\left(\ln{\s^2 \over m^2} -3\right)
\ee
where we have  introduced a renormalization scale $m$ and fixed counterterms by imposing the renormalization conditions
\bea
V[0] &=&0\\
{\pa^2 V [\s] \over \pa \s^2}_{|\s = m} &=& -{1\over \p f}.
\eea
It is then easy to see that, in order  for the effective potential to be independent of the scale $m$, $f$ has to satisfy (\ref{betaf}).
Taking $f$ to be the desired time-dependent source (\ref{flambda0}) and taking the scale $m$ to be the field $\s$ itself,  we obtain for the potential
\be
{1\over N} V[\s] = -{\s^2 \over 2 \p } \left(\ln{\s \cos t} +{3\over 2} \right)
\ee
As in the previous subsection, the two-derivative term in the effective action  is fixed by dimensional analysis to be of the form (\ref{effect}), leading to
the two-derivative effective action
 \be
 \G [\s ]  = - N \int d^2 x \left( V[\s ] +  c { \pa_\m \s \pa^\m \s \over 2 \s^2 }    +\ldots \right).\label{effectlambda0}
 \ee
This action allows for solutions of the form
\be
\s =-{\s_0 \over \cos t}\label{sigmalambda0}
\ee
 if we take  $c= \s_0^2(1-\half \ln \s_0^2)$. In view of (\ref{sigmaalpha}), this reproduces the behaviour of $\a$ in the bulk.

Going to Poincar\'e coordinates, the profile for $\a$ becomes
\be
\a = -{2\m \over 1 + y^\m y_\m}.\label{alphapoinc}
\ee
We note that the source term (\ref{dtracdef},\,\ref{flambda0}) was tuned precisely to make the 1-loop contribution to  the  effective potential (\ref{effectlambda0})  trivial.
We should therefore be able to find solutions of the classical theory (\ref{classferm}) with constant coupling $f$  where the bilinear $\bar \Psi^a \Psi_a$ has a profile of the form (\ref{alphapoinc}).  Indeed, such solutions have appeared in  the literature \cite{Akdeniz:1979wf}.

\section*{Acknowledgements}
We would like to  thank D. Anninos, A. Campoleoni, F. Denef, V. E. Didenko, G. Lucena G\'omez, T. Proch\'azka, P. Sundell and M. A. Vasiliev for useful comments and discussions.  This research was supported by the Grant Agency of the Czech Republic under the grant 14-31689S. The research of C. I. was also supported by the Russian Science Foundation grant 14-42-00047 in association with the Lebedev Physical Institute.

\bigskip

\appendix

\section{Spinor conventions}\label{App:conv}

We use conventions in which the $\mso(2,2)\simeq\mso(2,1)\oplus\mso(2,1)$ generators $M_{AB}$, $A,B = 0,1,2,0'$ satisfy the commutation relations
\be [M_{AB},M_{CD}]\ =\ 4\eta_{[C|[B}M_{A]|D]}\ ,\qquad
(M_{AB})^\dagger\ =\ M_{AB}\ ,\label{sogena}\ee
which can be decomposed, splitting $\eta_{AB}~=~(\eta_{ab};-1)$, $\eta_{ab}=\textrm{diag}(-1,+1,+1)$ with $a,b=0,1,2$, as
\be [M_{ab},M_{cd}]_\star\ =\ 4\eta_{[c|[b}M_{a]|d]}\ ,\qquad
[M_{ab},P_c]_\star\ =\ 2\eta_{c[b}P_{a]}\ ,\qquad [P_a,P_b]_\star\ =\
\l^{2} M_{ab}\ ,\label{sogenb}\ee
where $M_{ab}$ generate the Lorentz subalgebra $\mso(2,1)\simeq \msl(2,\RR)\simeq \msp(2,\RR)$, and $P_a=\l M_{0'a}$ with $\l$ being the inverse $AdS_3$ radius, related to the cosmological constant via $\L=-\l^{2}$. We set $\l=1$ in the body of the paper. In terms of the generators $J_a$, related to the $M_{ab}$ via
\bea M_{ab} \ = \ \epsilon_{ab}{}^cJ_c \ , \qquad J_a \ = \ -\frac{1}{2}\epsilon_{abc}M^{bc} \ , \ \ \qquad \epsilon^{012}=1  \ ,\eea
the $AdS_3$ isometry algebra reads
\be [J_{a},J_{b}]_\star\ =\ \epsilon_{ab}{}^cJ_c\ ,\qquad
[J_{a},P_b]_\star\ =\ \epsilon_{ab}{}^c P_{c}\ ,\qquad [P_a,P_b]_\star\ =\
\l^{2} \e_{ab}{}^cJ_c\ .\label{sogenc}\ee
One can use the Clifford element $\psi_1$, introduced in Section \ref{Sec:kinematics} to project onto the two $\msl(2,\RR)$ subalgebras of $\mso(2,2)$, to realize the translations as
\be P_a \ = \ J_a\psi_1 \ .\ee
In terms of the oscillators $y_\a$ introduced in \eq{yzcomm}, the generators can be realized as\footnote{This realization corresponds to the diagonal embedding of the Lorentz subalgebra into the $AdS_3$ isometry group $\mso(2,2)\simeq \mso(2,1)\oplus\mso(2,1)$ (see \emph{e.g.} \cite{Boulanger:2013naa} for a different, non-diagonal embedding).}
 \be
 J_{a}\ =\ \frac{i}8  (\s_{a})^{\a\b}y_\a\star y_\b ,\qquad P_{a}\ = \
 \frac{i}8 (\s_a)^{\a\b}y_\a \star y_\b\, \psi_1\ ,\label{mab}
 \ee
using van der Waerden symbols obeying
 \be
  (\s^{a})_{\a}{}^{\gamma}(\s^{b})_{\gamma}{}^{\b}~=~ \eta^{ab}\d_{\a}^{\b}\
 +\ (\s^{ab})_{\a}{}^{\b} \ ,\label{so4a}\ee
 \be ((\s^a)_{\a\b})^\dagger~=~ (\s^a)_{\a\b}~=~ (\s^a)_{\b\a} \ , \ee
 \be (\s^{ab})_{\a\b} \ = \ -(\s^{ba})_{\a\b} \ = \ \e^{abc} (\s_{c})_{\a\b} \ ,\ee
and raising and lowering spinor indices according to the
conventions
\bea
& v^\a \ = \ \e^{\a \b} v_\b\ , \qquad v_\b \ =  \ v^\a \e_{\a \b}  \ , \qquad  v_\a w_\b - v_\b w_\a \ = \ \e_{\a\b }v_\g w^\g \ ,&\\
& \e_{\a\b}\e^{\a\b} \ = \ 2 \ ,\qquad (\e_{\a\b})^\dagger \ = \ \e_{\a\b}\ . &
\eea
A convenient realization of the $(\s_{a})^{\a\b}$ is $\s^{a}=\{\id,\tilde{\sigma}_3,\tilde{\sigma}_1\}$, where $\tilde{\sigma}_a$ are the Pauli matrices. The map between vector and spinor indices is provided by the van der Waerden symbols as
\be x^{\a\b} \ = \ x^a(\s_{a})^{\a\b} \ , \qquad x^a \ = \ -\frac{1}{2}(\s^{a})_{\a\b}x^{\a\b} \ .\ee
The $\mso(2,2)$-valued connection
 \be
  \O~:=~\omega + e ~:=~  \frac12 \omega^{ab} M_{ab}+e^a P_a~:=~ -\frac{i}{4}
 \left( \omega^{\a\b}~y_\a \star y_\b
 +  e^{\a\b}~y_\a \star {y}_{\b}\psi_1\right)\
 ,\label{Omega}
 \ee
  \be
 \o^{\a\b}~=~ -\ft14(\s_{ab})^{\a\b}~\o^{ab}\ , \qquad \omega_{ab}~=~ -(\s_{ab})^{\a\b} \o_{\a\b}\ ,\ee
 \be e^{\a\b}~=~ -\ft{\l}{2}(\s_{a})^{\a \b}~e^{a}\ , \qquad e^a~=~ \l^{-1} (\s^a)_{\a\b} e^{\a\b}\ ,\label{convert}\ee
and its field strength
\be {\cal R}~:=~ d\O+\O\star \O~:=~ \frac12 {\cal R}^{ab}M_{ab}+T^a P_a ~:=~ \frac1{4i}
 \left({\cal R}^{\a\b}~y_\a \star y_\b
 +  T^{\a\b}~y_\a \star  y_{\b}\psi_1\right)\
 ,\label{calRdef}\ee
\be
 {\cal R}^{\a\b}\ =\ -\ft14(\s_{ab})^{\a\b}~{\cal R}^{ab}\ ,
 \qquad {\cal R}_{ab}~=~ - (\s_{ab})^{\a\b} {\cal R}_{\a\b}\ ,\ee
 \be
 T^{\a\b}\ =\ -\ft{\l}{2}(\s_{a})^{\a \b}~T^{a}\ ,
 \qquad T^a~=~ \l^{-1}  (\s^a)_{\a\b} T^{\a\b}\ .\ee
In these conventions, it follows that
 \be
 {\cal R}^{\a\b}~=~ d\o^{\a\b} -\o^{\a\g}\o_{\g}{}^{\b}-
 e^{\a\g} e_{\g}{}^{\b}\ ,\qquad
 T^{\a\b}~=~  de^{\a\b}-2 \o^{\g(\a}\wedge
 e_{\g}{}^{\b)}\ ,\ee
 \be
 {\cal R}^{ab}~=~ R^{ab}+\l^{2}
 e^a\wedge e^b\ ,\qquad R^{ab}~:=~d\o^{ab}+\o^a{}_c\wedge\o^{cb}\ ,\ee\be
T^a ~:=~d e^a+\o^a{}_b\wedge e^b\ ,
 \label{curvcomp} \ee
where $R_{ab}:=\frac12 e^c e^d R_{cd,ab}$ and $T_a:=e^b e^c T^a_{bc}$ are the Riemann and torsion two-forms.
The metric $g_{\mu\nu}:=e^a_\mu e^b_{\nu}\eta_{ab}$. The $AdS_3$ vacuum solution $\O_{0}=e_{0}+\o_{0}$ obeying $d\O_{0}+\O_{0}\star\O_{0}=0$, with Riemann tensor $ R_{0\m\n,\r\s}=
 -\l^{2} \left( g_{0\mu\rho} g_{0\nu\sigma}-
  g_{0\nu\rho} g_{0\mu\sigma} \right)$ and vanishing torsion, can be expressed as $\O_{0}=L^{-1}\star dL$ where the gauge function $L\in SO(2,2)/SO(2,1)$ \eq{Ly} and reads, in stereographic coordinates (see Appendix \ref{App:coset}),
 \be  e_{0}^a \ = \ \frac{2dx^a}{1-x^2} \ , \qquad e_{0}^{\a\b} \ = \ -\frac{dx^{\a\b}}{1-x^2} \ , \ee
 \be \o_{0}^{ab} \ = \ \frac{-4x^{[a}dx^{b]}}{1-x^2} \ , \qquad \o_{0}^{\a\b} \ = \ \frac{x^{\g(\a}dx_{\g}{}^{\b)}}{1-x^2} \ .  \ee

\section{Coset parametrization and stereographic coordinates}\label{App:coset}
$AdS_{d+1}$ can described as the following  hypersurface in $\RR^{d,2}$  (setting the $AdS$ radius to one, indices are lowered with $\h_{ab} \equiv {\rm \ diag} (-++ \ldots +)$):
\be
X^a X_a - (X^{0'})^2 = -1, \qquad a = 0, \ldots ,d
\ee
The stereographic coordinates $x^a$ used in the main text are given by
\bea
X^a &=& {2 x^a \over 1 - x^2}\\
X^{0'} &=& \pm \sqrt{1 + X^a X_a}\label{stereo}\\
ds^2 &=& {4 dx^adx_a \over (1- x^2)^2}
\eea

$AdS_{d+1}$ can also be viewed as the coset $SO(d,2)/SO(d,1)$. In the defining representation of $SO(d,2)$ the coset  and $SO(d,1)$ generators can be taken to be
\bea
P_a &=& E_{a, 0'} + (-1)^{\d_{a,0}} E_{ 0',a}  \\
M_{ab} &=&  E_{a, b} - (-1)^{\d_{a,0}} E_{ b,a} \qquad {\rm for\ a <b}
\eea
where the $E_{MN}$ are matrices with components $(E_{MN})_{mn}\equiv \d_{M,m} \d_{N,n}$.
Taking a canonical parametrization of the coset:
\be
L = e^{b^a P_a},
\ee
the embedding coordinates $X^a, X^{0'}$ are obtained by taking the last column of the matrix $L$ in the above representation \cite{Gilmore:2008zz}, leading to
\bea
X^a &=& { \sinh \sqrt{b^2} \over \sqrt{b^2}} b^a\\
X^{0'}  &=& \pm  \cosh \sqrt{b^2}
\eea
Comparing to (\ref{stereo}) we get the coset parametrization in stereographic coordinates:
\be
L \ = \ \exp { 2 {\rm arctanh} \sqrt{ x^2 } \over \sqrt{ x^2 }} x^a P_a.
\ee
Decomposing the left-invariant form $L^{-1} dL $ yields the vielbein and spin connection of $AdS_{d+1}$:
\bea
 L^{-1} dL &=& e^a P_a  + \half \o^{ab} M_{ab}\label{Ldecomp}\\
e^a &=& {2 dx^a \over 1- x^2}\\
\o^{ab} &=&  -{ 2 ( x^a dx^b- x^b dx^a)  \over 1- x^2}\\
ds^2 &=& (e, e) =  {4 dx^a dx_a \over (1- x^2)^2}
\eea

Specializing to $d=2$, and using the oscillator realization of the generators spelled out in Appendix \ref{App:conv}, %is (defining $J_a = \e_a^{\ bc} J_{bc}$):
%\bea
%J_0 &=& {1 \over 4} (T_{11} + T_{22}) = -{i \over 8} (y_1 \star y_1 + y_2 \star y_2)\\
%J_1 &=& {1 \over 4} (T_{11} - T_{22}) = -{i \over 8} (y_1 \star y_1 - y_2 \star y_2)\\
%J_2 &=& {1 \over 2} T_{12}  = -{i \over 8} (y_1 \star y_2 + y_2 \star y_1)\\
%P_a &=& J_a \psi_1
%\eea
the coset element reads
\bea
L &=& \exp_\star - {i \over 4}{ {\rm arctanh} \sqrt{ x^2 } \over \sqrt{ x^2 }} x^{\a \b} y_\a y_\b\psi_1 \ , \\
x^{\a \b} &=& \left( \begin{array}{cc} x^0+x^1 & x^2 \\ x^2 &   x^0-x^1 \end{array}\right) \ .
\eea

For completeness, we also give the definition of global $AdS_3$ coordinates $(\r, t, \varphi)$:
\be\begin{array}{lcl}
X^0 \ = \ \sin t \cosh r, &\qquad & X^1 \ = \ \sin \vf \sinh r\\
X^{0'} \ = \ \cos t \cosh r & \qquad & X^2 \ = \ \cos \vf \sinh r\label{global}
\end{array}\ee
and of Poincar\'e coordinates $(z, y^0, y^1)$:
\bea
X^\m &=& { y^\m \over z} \qquad \m = 0, 1\\
X^{2} &=& {z \over 2} \left( 1- { 1- y^\m y_\m \over z^2} \right)\\
X^{0'} &=& {z \over 2} \left( 1+ { 1+ y^\m y_\m \over z^2} \right)\label{Poincare}
%ds^2 &=& { d y^\m dy_\m + dz^2 \over z^2}
\eea

\section{Internal solution and deformed oscillators}\label{App:defosc}

Drawing on \cite{Prokushkin:1998bq,Sezgin:2005pv,Iazeolla:2007wt}, in this Appendix we shall recall the main steps involved in the solution of the internal equations \eq{internal1}-\eq{internal2}. Since $\wF'$ is also a spacetime constant, which we later choose to be $\wF'=\m\psi_2$, for notational simplicity in the following we shall denote $\n+\wF' =: \G$. Moreover, we shall also obtain the projector solutions mentioned in Section \ref{Sec:SS}.

We want to solve the deformed oscillator problem
\bea  [\wS'_\a,\wS'_\b]_\star \ = \ -2i\e_{\a\b}(1+\G \wK) \ , \qquad \{\wS'_\a,\wK\}_\star \ = \ 0 \ .\label{def}
\eea
Following \cite{Prokushkin:1998bq}, we start with a Lorentz-covariant Ansatz making use of the integral
representation
\bea \wS'_\a &=& \r\left[ \frac{1}{2}\,(y_\a+z_\a)\int_{-1}^1 dt~ n(t,k) \,e^{\frac{i}2(1+t)u}  \ - \  \frac{1}{2}\,(y_\a-z_\a)\int_{-1}^1 dt~ m(t,k) \,e^{\frac{i}2(1+t)u}\right]\ ,\label{ansatz}\eea
where $u:=y^\a z_\a$, which reduces \eq{def} to
\bea [\wS'_\a,\wS'_\b]_\star & =&   -2i\,\e_{\a\b}\,\int_{-1}^1 dt\int_{-1}^1 dt'~e^{\frac{i}2(1-tt')u}~n(t,k)m(t',k) \left\{1+\frac{i}{4}(1-tt')u\right\} \\[5pt]
& = & -2i\e_{\a\b}(1+\G \wK) \ .\eea
The rationale behind the choice of contour in the Laplace-transform-like Ansatz \eq{ansatz} is that it self-replicates under $\star$-product, in the sense that the latter maps $(t,t') \in [-1,1]$ into $-tt' \in [-1,1]$. One can then insert, in the l.h.s. of the last equation, $1 = \int_{-1}^1 ds\,\d(s-tt')$, turning it into the condition
\bea   -2i\,\e_{\a\b}\,\int_{-1}^1 dt\int_{-1}^1 dt'~e^{\frac{i}2(1-s)u}\,h(s,k)\left\{1+\frac{i}{4}(1-s)u\right\}
\ = \ -2i\e_{\a\b}(1+\G \wK) \ ,\label{step1}\eea
where we have set
\be  h(s,k) \ := \ n(t,k)\circ m(t',k)\ , \ee
and where $\circ$ defines the associative and commutative product \cite{Prokushkin:1998bq}
 \be
 (f\circ g)(s) \ := \ \int_{-1}^1 dt \int_{-1}^1 dt'
 \delta(s-tt')~f(t)~g(t')\ .
 \ee
Rewriting \eq{step1} as
\bea -2i\,\e_{\a\b}\,\int_{-1}^1 dt\int_{-1}^1 dt'~h(s,k)\left\{1-\frac{1}{2}(1-s)\frac{d}{ds}\right\}\,e^{\frac{i}2(1-s)u}
\ = \ -2i\e_{\a\b}(1+\G \wK) \ ,\label{step2}\eea
and integrating by parts, one gets
\bea  &&-i\e_{\a\b}\,\int_{-1}^1 ds~\{ h(s,k)+h'(s,k)(1-s)\} e^{\frac{i}2(1-s)u}\nonumber\\[5pt]
&+& i\e_{\a\b}\left[h(s,k)e^{\frac{i}{2}(1-s)u}(1-s)\right]^1_{-1}
\ = \ -2i\e_{\a\b}(1+\G \wK) \ ,\label{step3}\eea
where $h'(s,k) =\frac{d}{ds}h(s,k) $. If
\bea h(s,k)+h'(s,k)(1-s) \ = \ 2\d(s-1) \ ,\eea
the first term on the l.h.s. accounts for the $\G$-independent term on the r.h.s.
This condition can be satisfied with
\bea  h(s,k) \ = \ \d(s-1)-\frac{\G k}{2}(s-1) \ , \eea
which also ensures that the boundary term in the l.h.s. of \eq{step3} precisely reduces to the $\G$-deformation of the oscillator algebra.

Therefore, the Ansatz \eq{ansatz} reduces the deformed oscillator problem to the $\circ$-product problem
\be  (n(t,k)\circ m(t',k))(s)\ = \ \d(s-1)-\frac{\G k}{2}(s-1) \label{ringeq}\ee
Even and odd functions, denoted in the following by $f^\pm(t)$, are orthogonal with
respect to the $\circ$-product. Thus, the $\circ$-product problem splits into separate conditions for the even/odd parts of $n(t,k)$ and $m(t',k)$ (their arguments will be understood in the following):
\bea (n^{+}\circ m^{+})(s)&=&I^{+}_0(t)+\frac{\G k}2\ ,\label{nplus}\\[5pt]
(n^{-}\circ m^{-})(s)& = & I^{-}_0(t)-\frac{\G k}2 s\
,\label{nminus}\eea
where
\be I^{\pm}_0(t)\ =\
\frac12\left[\delta(t-1)\pm\delta(t+1)\right]\ .\label{I0}\ee
One proceeds \cite{Prokushkin:1998bq,Sezgin:2005pv,Iazeolla:2007wt} by writing
\bea n^\pm(t,k)&=& n_I(t,k)+n_p(t,k) \ = \ \sum_{\ell=0}^\infty \left(n_\ell I^\pm_\ell(t)+ \l_\ell p^\pm_\ell(t)\right)\
,\label{nexp} \eea %\\[5pt]
%m^\pm(t,k)&=& m_I(t,k)+m_p(t,k) \ = \ \sum_{\ell=0}^\infty \left(m_\ell I^\pm_\ell(t)+ \l'_\ell p^\pm_\ell(t)\right)\
%,\label{mexp}\eea
%
and analogously for $m^\pm$, where $I^{(\pm)}_0(t)$ (defined in \eq{I0}) and the
functions ($\ell\geq 1$)
 \bea
 I^{\s}_\ell(t)&=&\left[{\rm sign}(t)\right]^{\frac12(1-\s
 )}~\int_{-1}^1 ds_1 \cdots \int_{-1}^1 ds_\ell~\delta(t-s_1\cdots
 s_\ell)\nonumber\\[5pt]
 &=&\left[{\rm sign}(t)\right]^{\frac12(1-\s)}{\left(\log
 \frac1{t^2}\right)^{\ell-1}\over (\ell-1)!}\ ,
 \eea
obey the algebra ($\ell_1,\ell_2 \geq 0$)
 \be
 I^{\s}_{\ell_1}\circ I^{\s}_{\ell_2}\ =\ I^{\s}_{\ell_1+\ell_2}\ ,
 \label{ring}
 \ee
and where $p^\s_\ell(t)$ ($\ell\geq 0$) are the $\circ$-product projectors
\bea p^{\s}_\ell(t)&=& {(-1)^\ell\over \ell!} \d^{(\ell)}(t)\ ,\qquad \s\ =\
(-1)^\ell\ ,\label{pk}\eea
that obey
\bea p^\s_\ell\circ f&=& L_\ell[f] p^\s_\ell\ ,\qquad L_\ell[f]\ =\ \int_{-1}^1
dt~ t^\ell f(t)\ .\label{proj1}\eea
In particular,
\bea p^{\s}_{\ell_1}\circ p^{\s}_{\ell_2}&=& \delta_{\ell_1\ell_2}p^{\s}_{\ell_1}\
.\label{proj2}\eea
Substituting the expansion \eq{nexp} into \eq{nplus} and
\eq{nminus}, one finds, in view of \eq{ring}, \eq{proj1} and
\eq{proj2}, manageable algebraic equations. In particular, the property \eq{ring} makes it possible to map the $\circ$-product between any two function expanded over the $I^{\s}_k$ into the ordinary product of two ordinary functions of the variable $\xi$. Therefore, substituting to any $f^\s(t)=\sum_{k=0}^\infty f_\ell I^\s_\ell(t)$ its symbol $\tf(\xi)=\sum_{\ell=0}^\infty f_\ell \xi^\ell$, one has
\bea \widetilde{(f\circ g)} (\xi) \ = \ \tf(\xi)\,\tg(\xi) \ . \label{symb}\eea

We shall exploit the mapping to the symbols in order to solve \eq{nplus} and \eq{nminus}. After that, we shall show that any so-obtained $\G$-dependent solution (including the undeformed oscillators $\wS'_\a=\rho z_\a$) can be ``dressed up'' with projector solutions that bring in new, discrete moduli. So in the following, and until the projectors will be reintroduced,  we shall set $\l_\ell=\l'_\ell=0$ and drop the $I$ index used in \eq{nexp}.

By virtue of \eq{symb}, \eq{nplus} and \eq{nminus} are mapped to%\footnote{By the property \eq{ring}, $\xi^k$ is the symbol of $I_k$. Thus, the rationale behind the deformation term in \eq{nminussymb} is that the function $f(t)=t$ admits the expansion $t=\sum_{k=1}^\infty (-\frac{1}{2})^{k-1}I_k^{(-)}(t)$.}
\bea \tn^{+}\tm^{+}&=& 1+\frac{\G k}2 \xi\ ,\label{nplussymb}\\[5pt]
\tn^{-} \tm^{-} & = &1-\frac{\G k}2 \frac{\xi}{1+\xi/2}\
.\label{nminussymb} \eea
Note that, differently from the four-dimensional minimal-bosonic model solutions of \cite{Sezgin:2005pv,Iazeolla:2007wt}, the above equations constrain the product of two functions, and therefore admit a class of solutions parameterized by an arbitrary function. However, it is only the symmetric solution $n(t,k)=m(t,k)$ that satisfies the reality condition on $S_\a$ \eq{real}, and for this reason we shall focus on it in the following.
%(
%\bea\tn^{+}(\xi,k) &=& f(\xi,k) \sqrt{1+\frac{\G k}2 \xi}\ ,\ \quad \tm^{+} (\xi,k) \ = \ f^{-1}(\xi,k) \sqrt{1-\frac{\nu k}2 \xi}\ ,\label{nmplus}\\[5pt]
%\tn^{-}(\xi,k) &=& g(\xi,k) \sqrt{1+\frac{\nu k}2 \frac{\xi}{1+\xi/2}}\ ,\ \quad \tm^{-} (\xi,k) \ = \ g^{-1}(\xi,k) \sqrt{1-\frac{\nu k}2 \frac{\xi}{1+\xi/2}}\ .\label{nmminus}\ \eea
%
After solving for $\tn^{\pm}=\tm^{\pm}$, one can transform back, and thus reconstruct $n(t,k)$. Taking into account the initial condition $\wS'_\a\left|_{\G=0}\right. = \rho z_\a$ implies that $n(t,k)\left|_{\G=0}\right. =\d(t+1)$, which fixes $n = n^+-n^-$.

%We shall now study the two extreme cases of $f(\xi,k) = \sqrt{1+\frac{\nu k}2 \xi}$, $ g(\xi,k) = \sqrt{1+\frac{\nu k}2 \frac{\xi}{1+\xi/2}}$ (or the opposite choice $f^{-1}(\xi,k) = \sqrt{1+\frac{\nu k}2 \xi}$, $ g^{-1}(\xi,k) = \sqrt{1+\frac{\nu k}2 \frac{\xi}{1+\xi/2}}$) and $f=g=1$, which we may refer to, respectively, as ``most asymmetric'' and ``symmetric'' solution of the $\circ$-product problem, as well as the ``mixed'' solutions $f(\xi,k) = \sqrt{1+\frac{\nu k}2 \xi}$, $ g^{-1}(\xi,k) = \sqrt{1+\frac{\nu k}2 \frac{\xi}{1+\xi/2}}$ and $f^{-1}(\xi,k) = \sqrt{1+\frac{\nu k}2 \xi}$, $ g(\xi,k) = \sqrt{1+\frac{\nu k}2 \frac{\xi}{1+\xi/2}}$.

\paragraph{\emph{Symmetric solutions.}}

Let us first examine the even sector.
\bea \tn^{+}(\xi,k) &=& \tm^{+} (\xi,k) \ = \ \sqrt{1+\frac{\G k}2 \xi}\ . \eea
Expanding in power series of $\xi$ and transforming back to the corresponding $t$-functions one
finds
\bea n^{+}(t,k) &=& m^{+} (t,k) \ = \ I_0^+(t) + \sum_{\ell=0}^\infty {1/2  \choose \ell+1} \left(\frac{\G k}{2}\right)^{\ell+1}I^+_{\ell+1}(t) \\[5pt] %&=&  I_0^+(t) -\frac{1}{2} \sum_{n=0}^\infty \frac{(1/2)_n}{(2)_n} \left(\frac{\nu k}{2}\right)^{n+1}\frac{(-\log t^2)^n}{n!} \\
&=& I_0^+(t) +{\G k\over 4} ~{}_1\!
F_1\left[\frac12;\,2;\,\log |t|^{+\G k}\right]\ .\label{nmsym}\eea

Getting an expansion in $I^-_\ell(t)$ is more complicated in the odd sector. Resorting to the trick used in \cite{Sezgin:2005pv} to transform back %that makes use of the Laplace transform of (the non-trivial part of)
\bea \tn^{-}(\xi,k) &=&  \tm^{-} (\xi,k) \ = \ \sqrt{1-\frac{\G k}2 \frac{\xi}{1+\xi/2}}\ .\label{nmminus}\ \eea
%
%Let's begin by introducing the function
%
%\bea q(t,k) \ := \ \textrm{sign}(t)\,\hat{q}\left(\log (1/t^2)\right) \ = \  \textrm{sign}(t)\,\sum_{n=1}^\infty \hat{q}_k I^+_k(t)\eea
%
%corresponding to the non-trivial part of $n^-(t,-k)=m^-(t,k)$. Its symbol is $\tilde{\hq}(\xi,k)=\textrm{sign}(t)\,\sum_{n=1}^\infty \hat{q}_k \xi^k$. We proceed by introducing (we shall in the end set $\z=-\log t^2$)
%
%\bea \hQ \ := \ \int_0^\z d\z'\,\hq(\z',k) \ = \ \sum \hat{q}_k \frac{\z^k}{k!} \ ,\eea
%
%and by using $\xi^k=\int_0^\infty d\z\, e^{-\z}(\xi\z)^k/k!$, we get
%
%\bea \tilde{\hq}(\xi,k) \ = \ \int_0^\infty d\z\, e^{-\z} \hQ(\xi\z) \ .\eea
%
%This Laplace-transform can be inverted to give
%
%\bea \hQ(\z) \ = \ \int_{\g-i\infty}^{\g+i\infty}\frac{dz}{2\pi i z}\,e^{\z z}\,\tilde{\hq}\left(\frac{1}{z},k\right) \ ,\eea
%
%with  $ \g > \textrm{max} \{Re{z_i} : z_i \,\textrm{pole or branch cut of} \, \frac{1}{z}\tilde{\hq}\left(\frac{1}{z},k\right) \}$.
to the $t$-functions, one obtains
%
%\bea \hat{q}\left(\log (1/t^2)\right) & = & \frac{d}{d\z} \hQ(\z)\left|_{\z=-\log t^2}\right. \ = \ \int_{\g-i\infty}^{\g+i\infty}\frac{dz}{2\pi i}\,e^{\z z}\,\left[\sqrt{1-\frac{\n k}{1+2z}}-1\right] \bigg{|}_{\z=-\log t^2}\\[5pt]
 %& = & -\frac{\nu k|t|}{4}~{}_1\!
%F_1\left[\frac12;\,2;\,\log |t|^{-\nu k}\right] \ ,\eea
%
%and therefore
%
\bea n^{-}(t,k) &=& m^{-} (t,k) \ = \ I_0^-(t) -{\G k\over 4} \,t~{}_1\!
F_1\left[\frac12;\,2;\,\log |t|^{-\G k}\right]\ .\label{nmsymapp}\eea

Imposing the initial conditions as explained above, one finally obtains
\bea && n(t,k) \ = \ m(t,k) \nonumber \\[5pt]
& = &\ \d(t+1) +  {\G k\over 4} \left({}_1\!
F_1\left[\frac12;\,2;\,\log |t|^{\G k}\right]+t{}_1\!
F_1\left[\frac12;\,2;\,\log |t|^{-\G k}\right]\right) \ ,\label{nsym}\eea
that is,
\bea && \ \ \ \ \wS'_\a \ = \ \rho z_\a\nonumber \\[5pt]
&+& \rho z_\a{\G k\over 4}\int_{-1}^1 dt~e^{\frac{i}2(1+t)u}\left({}_1\!
F_1\left[\frac12;\,2;\,\log |t|^{\G k}\right]+t{}_1\!
F_1\left[\frac12;\,2;\,\log |t|^{-\G k}\right]\right) \label{Sreg}\ ,\eea
which corresponds to \eq{S0reg} or to \eq{primesol} for $\G=\n$ or $\G=\n+\m\psi_2$, respectively.

\paragraph{\emph{Dressing with projectors.}}  Let us now show how the above-obtained solutions can be dressed-up with projectors. To begin with, we now plug the full expansion \eq{nexp} into \eq{nplus}-\eq{nminus}, with  $n_I,\,m_I$ correspond to the symmetric solution above presented. This means that, taking into account the property  \eq{proj1}, the coefficients $\l_\ell,\,\l'_\ell$ must satisfy
\bea \l_\ell L_\ell[m^\pm_I(t,k)]+\l'_\ell L_\ell[n^\pm_I(t,k)]+\l_\ell \l'_\ell \ = \ 0 \ . \label{lambda}\eea

For symmetric solutions $n(t,k)=m(t,k)$, which implies
\bea \l_\ell&=& \l'_\ell \ = \ - 2\th_\ell L_\ell[n^\pm_I]\ ,\qquad \th_\ell\in\{0,1\}\
.\label{lambdak}\eea
i.e., recalling the definition of $L_\ell$ from \eq{proj1},
\bea \l_{2n} & = & -2\th_{2n}\sqrt{1+\frac{\G k}{2n+1}} \ , \\[5pt]
 \l_{2n+1} & = & -2\th_{2n+1}\sqrt{1-\frac{\G k}{2n+3}} \ . \eea
The resulting $\wS'_\a$ is therefore a sum of the ``regular'' part of the expansion \eq{Sreg} and of the projector part, which can be written as
\bea
\wS'_\a & = & \wS_\a^{\prime reg} + \wS_\a^{\prime proj} \eea
where
\bea &\wS_\a^{\prime proj} & \ = \ - 2\rho z_\a\int_{-1}^1 dt ~e^{\frac{i}2(1+t)u}\quad \times\nonumber\\[5pt]
&\times& \sum_{n=0}^\infty  \th_n \left[\frac{1+(-1)^n}{2}\sqrt{1+\frac{\G k}{1+n}}-\frac{1-(-1)^n}{2}\sqrt{1-\frac{\G k}{2+n}}\right] p_n(t)  \label{typroj1}\ .
\eea
Performing the $t$-integrals gives
\bea
 \!\!\!\!\!\! \!\!\!\!\!\! \wS_\a^{\prime proj}   =  - 2\rho z_\a\sum_{n=0}^\infty  \th_n (-1)^n P_n(u) \left[\frac{1+(-1)^n}{2}\sqrt{1+\frac{\G k}{1+n}}-\frac{1-(-1)^n}{2}\sqrt{1-\frac{\G k}{2+n}}\right] \ ,\label{typroj2}
 \eea
where
\bea P_n(u)&=&  {1\over
n!}\left({-iu\over 2}\right)^n e^{\ft{iu}2}\ ,\label{Pn1}\eea
are projectors in the $\star$-product algebra given by functions of
$u$, \emph{viz.}
\bea P_n\star P_m\ =\
\delta_{nm}P_n\ .\eea
Let us define now the combinations $A_\a:= \frac{1}{2}(y_\a+z_\a)$ and $ A^\dagger_\a  := \frac{1}{2i}(y_\a-z_\a)$. They span the complexified Heisenberg algebra  $[A_\a,A^{\dagger\,\b}]_\star \ = \ \d_\a^\b$.  Recalling now that the $\star$-product law \eq{starprod} corresponds to normal-ordering with respect to $A_\a$ and $A^\dagger_\a$, it becomes clear that the $P_k(u)$ are the Fock-space projectors
\bea P_n(u) \ = \ \ket{n}\bra{n} \ = \ {1\over
n!} A^{\dagger\,\a_1}\star\ldots \star A^{\dagger\,\a_n}\star e^{-N}\star A_{\a_1}\ldots \star A_{\a_n} \ , \eea
where $N$ is the number operator $A^{\dagger\,\a}\star A_\a = A^{\dagger\,\a} A_\a$, and, in normal ordering, $e^{-N} = P_0$.

Note that, as first observed in \cite{Iazeolla:2007wt} for the four-dimensional case, the projector part of the solution is non-trivial even at $\G=0$, i.e., the projectors give rise to seemingly new vacuum solutions. In particular, for $\G=0$,
\bea \wS'_\a \ = \ \rho z_\a \left(1-2\sum_{n=0}^\infty \th_nP_n(u)\right) \ , \eea
and in this sense, as explained in Section \ref{Sec:SS}, the projector-dependent term above may be regarded as a flat yet non-trivial $Z$-space connection. Note also that, interestingly, the projector solutions \eq{typroj2} exhibit a regular behaviour for $\G$ in a finite region around the origin, with branching points occurring at the critical values $\G = 2n+1$, $n\in \mathbb{Z}$, at which the higher-spin algebra develops infinite-dimensional ideals \cite{Feigin88,Vasiliev:1997dq,Prokushkin:1998bq,Vasiliev:1999ba}. We defer a study of such solutions to future work.

\section{Internal solution and $\star$-product lemmas}\label{lemmas}

In this Appendix we collect a few technical remarks that are useful in obtaining the results of Section \ref{secgauge}. In particular, starting from the expression \eq{primesol2}, the deformation of the Lorentz generator in the ``nothing gauge'' can be written as
\bea {1 \over 2} \{\widehat S'_\a,\widehat S'_\b \}_{\star} & = & z_\a z_\b
+ \m \,z_\a z_\b \int_{-1}^1 \,ds \,(1-s) \left[(1-s)F^-(\m\ln|s|)e^{\frac{i}{2}(1+s)u}\right.\nonumber\\
&-&\left.(1+s)F^+(\m\ln|s|)\,e^{\frac{i}{2}(1-s)u}\,k\psi_2\right]\nonumber\\
&-&\mu^2  \int_{-1}^1 \,ds\int_{-1}^1 \,d\tilde s\,(1-s)(1-\tilde s)  \left\{\left[F^+(\m\ln|s|)F^+(\m\ln|\tilde s|)(a^\dagger_\a-\tilde s a_\a)(a_\b-sa^\dagger_\b)\right.\right.\nonumber\\
 &&+ \left.\left. F^-(\m\ln|s|)F^-(\m\ln|\tilde s|)(a^\dagger_\a+\tilde s a_\a)(a_\b+sa^\dagger_\b)\right]e^{\frac{i}{2}(1-s\tilde s)u}\right.\nonumber\\
 &+&\left[F^+(\m\ln|s|)F^-(\m\ln|\tilde s|)(a^\dagger_\a+\tilde s a_\a)(a_\b-sa^\dagger_\b) \right.\nonumber\\
&&+ \left.\left. F^-(\m\ln|s|)F^+(\m\ln|\tilde s|)(a^\dagger_\a-\tilde s a_\a)(a_\b+sa^\dagger_\b)\right]e^{\frac{i}{2}(1+s\tilde s)u}\,k\psi_2 \right\} \ ,\label{S'2}\eea
where we have introduced the mutually conjugated oscillators $a_\a:=(y_\a+z_\a)/2$, $a^\dagger_\a:=(y_\a-z_\a)/2$, with $[a_\a,a^\dagger_\b]_\star= i\e_{\a\b}$. This expression contains bilinears times exponentials in $(y,z)$, with or without an insertion of $\psi_2$. In order to evaluate the $\star$-products in \eq{LrotS}, it is useful to represent the terms in \eq{S'2} with the help of ``sources'' $\r_\a$ and $\r'_\a$, e.g.,
\bea L^{-1}\star y_\a z_\b \, e^{itu}\,\psi_2^A\star L \ = \ \left.\frac{\partial}{\partial \r^\a
}\frac{\partial}{\partial \r^{\prime\b}} L^{-1}\star e^{itu+\r^\g y_\g+ \r^{\prime\g} z_\g}\,\psi_2^A \star L\right|_{\r=\r'=0} \ . \eea
where $A=\{0,1\}$. By virtue of \eq{psialg}, the $L$-rotation in \eq{LrotS} acts differently on the coefficients of the identity and of $\psi_2$ above, giving rise to two types of structures. Using the matrix notation, e.g., $MaM : = M^\a a_\a{}^\b M_\b $, the latter can be evaluated as
\bea
&&L^{-1}\star e^{itu+\r y+ \r' z}\star L
\ = \ \frac{1-a^2}{1-a^2(1-2s)^2}\,\exp \bigg[itu+\r y+ \r' z \nonumber\\
&-&\left.\frac{i}{1-a^2(1-2s)^2}\left(\half\psi_1(MaM-NaN)-a^2(1-2s)MN\right)\right] \ ,
\eea
where we defined the spinors
\bea
M^\a &:= & y^\a(1-t)-t z^\a-i(\r^\a-\r^{\prime\a}) \ , \\
N^\a & := & y^\a(1-t)+t z^\a+i(\r^\a+\r^{\prime\a})  \ ,
\eea
and the remaining variables are defined in the body of the paper; and, with the same definitions,
\bea
&& \psi_2 L\star e^{itu+\r y+ \r' z}\star L \,
\ = \ \psi_2 \frac{1-a^2}{1+a^2(1-2s)^2}\,\exp \bigg[itu+\r y+ \r' z \nonumber\\
&+&\left.\frac{i}{1+a^2(1-2s)^2}\left(\half\psi_1(MaM-NaN)-a^2(1-2s)MN\right)\right] \ .
\eea
Using these lemmas for the $L$-rotation of the various terms in \eq{S'2} and then projecting onto $z=0$ one obtains the results in \eq{SSphys} and \eq{SStwist}.

\section{Characters of $SL(2,\RR)$ $\subset$ $HS [\l ]$}\label{apptraces}
In this Appendix we compute the trace  of an exponential of a quadratic expression in the deformed oscillators:
\bea
\tr g  &\equiv&\tr \exp_\star \left( -{i \over 8} v^{\a\b}\tilde  y_\a \star \tilde y_\b \right)\\
&=& \tr \exp_\star \left(- v^a J_a \right) \label{charsl2}
\eea
where the trace is normalized such that $\tr \exp_\star 0 =1$.
Such traces can be seen as  characters
of  the  $SL(2,\RR)$ subgroup of  the higher spin gauge group $HS[\l]$.  The character is to be taken in the representation where  the quadratic Casimir takes the value
\be
C_2 = - J_0^2 + J_1^2 + J_2^2 = {1 \over 4} (\l^2-1)
\ee
where $\l = \half (1 - \n k)$. Hence the character can be computed by taking the character in the $N$-dimensional representation of $SL(2,\RR)$
and then continuing $N \to \l$.

The character in the $N$-dimensional representation of $SL(2,\RR)$ can be obtained by analytic continuation from the
character in the $N$-dimensional representation of $SU(2)$. For this we have the standard expression (see e.g. \cite{Biedenharn:1981er}, p.144)
\be
\tr_N \exp \left(- i \o \vec{n} \cdot \vec{T} \right)= {\sin {N \o \over 2} \over N \sin {\o \over 2}}.
\ee
where $\vec{n} \cdot \vec{n} = 1$ and $T_i$ are  the $SU(2)$ generators in the $N$ dimensional representation with commutation relations
$[T_i, T_j ] = i \e_{ijk} T_k$.

To continue this expression to the  $N$-dimensional representation of $SL(2,\RR)$, we can take the $SL(2,\RR)$ generators to be $T_1 = - i J_0, T_2 =J_1, T_3 = J_2$, so that in order to compute the character (\ref{charsl2}), we have to take $\o = i \sqrt{v^2}$. Continuing $N \to \l$ then leads to
\be \tr g  = {\sinh {\l \sqrt{v^2} \over 2} \over \l \sinh {\sqrt{v^2}\over 2}}.\ee
We note also that $\tr g = \tr g^{-1}$, so that we arrive at the formula needed in the main text:
\be
\tr \cosh_\star \left( -{i \over 8} v^{\a\b}\tilde  y_\a \star \tilde y_\b \right)= {\sinh {\l \sqrt{v^2} \over 2} \over \l \sinh {\sqrt{v^2}\over 2}}. \label{charsl22}
\ee

\section{From Fubini to Liouville instantons}\label{appfubini}
In this Appendix we show that the solutions (\ref{sigmasolpoinc}) of classical Liouville theory with a negative potential can be seen as a $d\to 2$ scaling limit of the
well-known Fubini instantons \cite{Fubini:1976jm} which exist in scalar theories in $d>2$.
We consider, in dimension $d > 2$, and in Minkowski signature for definiteness, the following action for a  scalar field $\f$:
\be
S_d = - \int d^d y \left[  \half (\pa \f)^2 + g \f^{ 2d \over d-2} \right] .
\ee
The coupling $g$ is dimensionless and these are classically conformally invariant theories. When $g<0$, the $\f=0$ vacuum
is unstable and the theory admits Fubini instantons \cite{Fubini:1976jm} with $O(d-1,1)$  (or  $O(d)$ in Euclidean signature) symmetry:
\be
\f = \left( {{d-2\over \sqrt{- 2 g}} { \r \over \r^2 + y^\m y_\m } } \right)^{d-2 \over 2}\label{Fubiniinst}
\ee
where $\r >0$ is the size of the instanton.

When $d=2$, the above action breaks down, while the Fubini solution behaves as
\be
\f = 1 + {d-2 \over 2}\ln {(d-2) \r \over \sqrt{-2 g} (\r^2 + y^\m y_\m ) } +\calo (d-1)^2.
\ee
This suggests the following  scaling limit under which the solution
stays regular. We  define
\be
\f = 1 + {d-2\over 2}\left( {\g \psi \over 2} + \ln {  (d-2) \g \over 4} \right).
\ee
for some constant $\g$. In the limit $d\to 2 $, the leading part of the action is  of the Liouville type:
\be
S \sim - \left({\g (d-2)\over 4}\right)^2 \int d^2 x \left[  \half (\pa \psi)^2 + g e^{ \g \psi} \right]+  \calo (d-2)^3
\ee
and the Fubini solution (\ref{Fubiniinst}) reduces to the $O(1,1)$   (or $O(2)$) invariant solution (\ref{sigmasolpoinc}) of Liouville theory with negative potential:
\be
\psi =  {2 \over \g} \ln {4 \r \over \sqrt{-2 g} \g ( \r^2 + y^\m y_\m )} .
\ee

\end{document}